\documentclass[journal=jacsat]{achemso}
\setkeys{aps}{}

\pdfoutput=1
\usepackage[version=3]{mhchem} 
\usepackage{color}
\usepackage{mathrsfs}
\usepackage{graphicx, subfigure}
\usepackage{bm}

\newcommand{\vs}[1]{\boldsymbol{#1}}                      
\newcommand{\abs}[1]{\left| #1 \right|}
\newcommand{\onlinecite}[1]{\hspace{-1 ex} \nocite{#1}\citenum{#1}} 
\SectionNumbersOn

\author{D. Fiocco}
\affiliation[Ecole Polytechnique F\'ed\'erale de Lausanne (EPFL)]{Institute of Theoretical Physics (ITP), Ecole Polytechnique F\'ed\'erale de Lausanne (EPFL), CH-1015 Lausanne, Switzerland}
\author{G. Pastore}
\affiliation[Dipartimento di Fisica dell'Universit\`a di Trieste and CNR-IOM]{%
Dipartimento di Fisica dell'Universit\`a di Trieste and CNR-IOM, Strada Costiera 11, 34151 Trieste, Italy
}
\author{G. Foffi}
\email{giuseppe.foffi@epfl.ch}
\affiliation[Ecole Polytechnique F\'ed\'erale de Lausanne (EPFL)]{Institute of Theoretical Physics (ITP), Ecole Polytechnique F\'ed\'erale de Lausanne (EPFL), CH-1015 Lausanne, Switzerland}

\title{Effective forces in square well \\ and square shoulder fluids}

\begin{document}

\begin{abstract}
We derive an analytical expression for the effective force between a pair of macrospheres immersed in a sea of microspheres, in the case where the interaction between the two unlike species is assumed to be a square well or a square shoulder of given range and depth (or height).
This formula extends a similar one developed in the case of hard core interactions only.
Qualitative features of such effective force and the resulting phase diagram are then analyzed in the limit of no interaction between the small particles. 
Approximate force profiles are then obtained by means of integral equation theories (PY and HNC) combined with the superposition approximation and compared with exact ones from direct Monte Carlo simulations.
\end{abstract}

\section{\label{sec:Intro}Introduction}

Colloidal systems are ubiquitous and have attracted a growing interest in the last decades. 
The impact on everyday life is vast and ranges from food \cite{mezzenga2005understanding} to materials, from biology to photonic crystals \cite{evans1994colloidal}.
On a more fundamental level, experiments on colloids have been the testing ground for several physical phenomena, from thermodynamics \cite{pusey1986phase, gast1983polymer} to the glass transition \cite{foffi2002phase}. In general, the interest comes from the possibility of tuning macroscopic properties by encoding them at the level of the interactions between the constituents of colloidal solutions.
For these reasons these systems have dominated the scientific discussions in these fields during the last years. In particular, the possibility of engineering systems with tunable microscopic properties allowed the observation of exotic phenomena like the Alder transition for hard spheres \cite{alder1957phase}, the metastable liquid-liquid phase separation \cite{gast1998simple}, the existence of two distinct glassy phases \cite{pham2002multiple, dawson2000higher} for short ranged attractive potentials and the emergence of a thermodynamically stable cluster phase  \cite{stradner2004equilibrium, sciortino2004equilibrium}. \\
As most soft-matter systems, colloids are characterized by a large number of degrees of freedom spanning several length and time scales. In several circumstances, however, it is possible to integrate out some of these degrees of freedom to gain more insight. In fact, colloids are one of the prototypical systems where a coarse-grained approach can be extremely fruitful. Typically, the extra degrees of freedom are mapped into some effective interaction among colloidal particles \cite{likos2001effective}, enabling the use of the full arsenal of statistical mechanics. In particular, most of the results for simple liquids (integral equations, perturbation theory, mode coupling theory, etc.) can be successfully used to address questions of great relevance for colloidal systems. \\
A classical example of such success is the case of depletion interactions. In the case of a binary mixture of colloidal hard spheres, when the diameters of the two species are very different, an entropic force starts to set in \cite{frenkel1999entropy}. This results in a net attraction between the colloids belonging to the largest species. Asakura and Oosawa \cite{asakura1954interaction} and, independently, Vrij \cite{vrij1976polymers} derived an effective interaction potential casting the problem of a binary mixture into an effective single-component system. This very simple result was the beginning of a series of investigations (see for instance (\onlinecite{cinacchi2007large, lajovic2009depletion})). It was possible, for the first time, to investigate a system of particles interacting with an attraction whose range could be tuned, something that is not feasible for atomic or molecular systems. The profound effects both on thermodynamics and the dynamics of a short range attraction are now well established.  \\
 The use of effective interactions, far from being restricted to colloidal systems, has important implications also for proteins, especially for what concerns crystal nucleation and phase behavior. By modeling proteins as short range attractive colloids it has been possible, for example,  to rationalize the enhancement of crystal nucleation in the proximity of the liquid-liquid critical point \cite{wolde1997enhancement}. This extremely simplified modeling has been also useful in the case of binary mixtures of eye-lens proteins, whose experimental behavior was successfully modeled using a simple hard sphere potential with specific square well attractions \cite{stradner2007new, dorsaz2008colloidal}. Here, a fundamental role was played by the intensity of the interspecies attraction: by varying it, it was possible to reduce the instability with respect to demixing due to depletion.
 Apart from the medical and biological implications, this work showed how a modification of the mutual interaction among the components can alter the phenomenology of the whole mixture. \\
 The aim of this paper is to rationalize this fact in terms of effective interactions. We investigated the effective interactions between the larger component of an asymmetric binary mixture of hard spheres where the potential between the two components has the form of a square well or a square shoulder. To derive our results we followed a reasoning similar to Attard's derivation of the effective \textit{force} between two hard spheres immersed in a sea of smaller ones \cite{attard1989spherically}.
We  generalized Attard's argument to the case of square well (or shoulder) interspecies interaction. The calculation of the actual force, as in (\onlinecite{attard1989spherically}), requires the evaluation of the density profile of the smaller spheres around a pair of larger ones. We performed this step following different methods.
We shall first propose a simple approximation that treats the smaller component as an ideal gas. In this way, we shall find a generalization of the Asakura-Oosawa (AO) analytical expression of the force to the case in which the mutual interaction potential is an attractive square well or a repulsive shoulder. Within such approximation we'll explore, by means of first order thermodynamic perturbation theory, the qualitative changes in the phase diagram determined by the mutual interaction. The analogy with the case of binary mixtures of eye-lens proteins are evident.
As for the purely repulsive case, the approximation above holds only when the interactions between the smaller particles are negligible. Thus we improved the results of the force calculation using an integral equation approach combined with the superposition approximation for the case of hard sphere interaction between the smaller species. All the aforementioned theoretical results have been tested with respect to the ``exact values'' of the force as calculated directly from new Monte Carlo simulations.\\
The paper is organized as follows:
in  \ref{sec:ForceDerivation} we present our generalization of Attard's derivation of the effective force in the case of square well/shoulder interactions between the species of a binary mixture;
in  \ref{sec:DensityDetermination} the methods used to estimate the density profile needed to compute the effective force are described. Details about the parameters used are also provided and a description of the qualitative effect of the effective interactions on the phase diagram is given;
in  \ref{sec:Results} quantitative results obtained with the different methods are shown and discussed. Finally in  \ref{sec:Conclusion} conclusions of this work are drawn.

\section{\label{sec:ForceDerivation}Force in the square well/shoulder case}

\begin{figure}
	\includegraphics[width = 0.6 \textwidth]{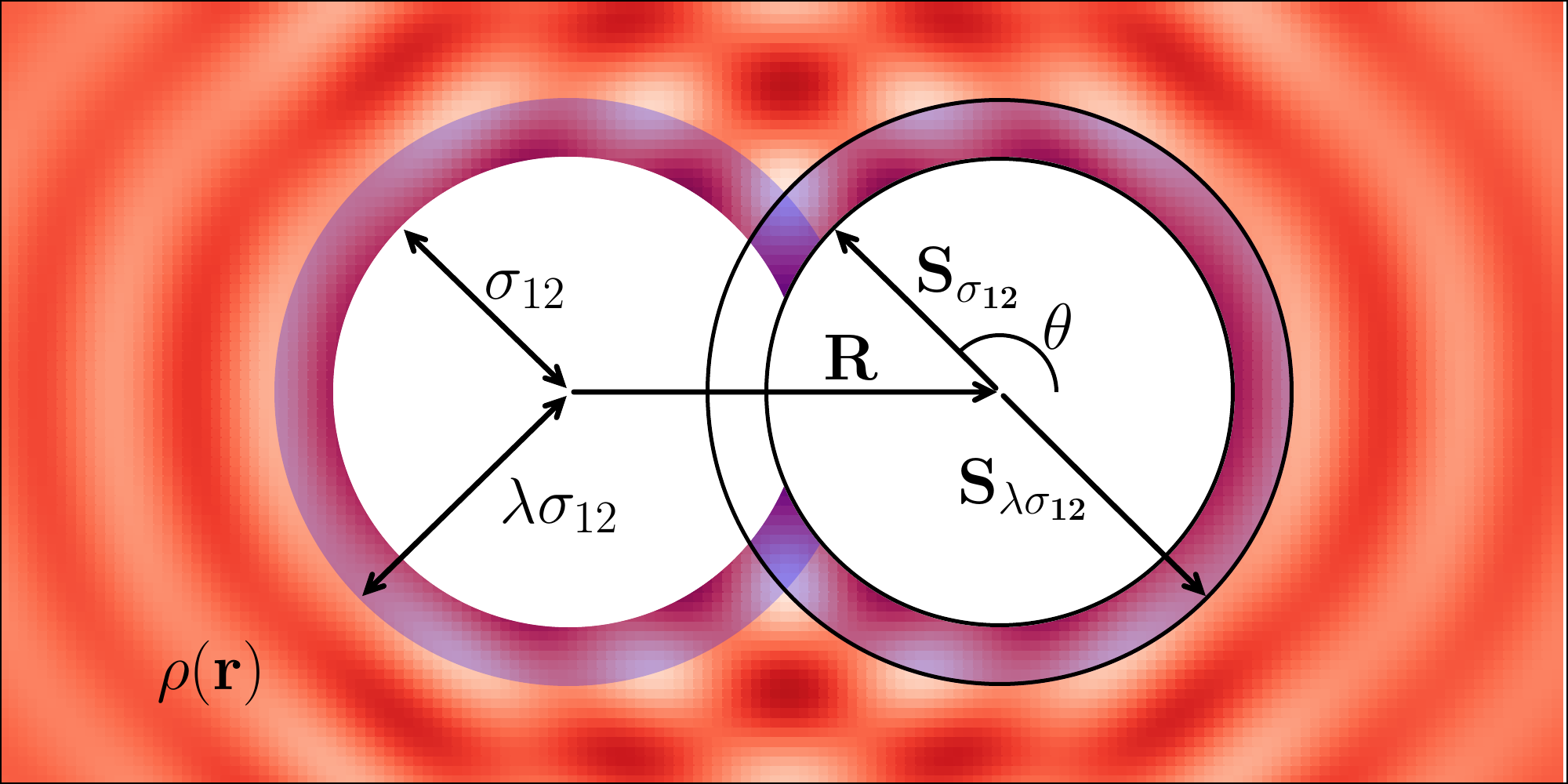}
	\caption{\label{fig:Parameters} Scheme of the geometrical parameters that appear in  \ref{eq:AttardForce} and \ref{eq:FinalForce}. The solid white area indicates the region inaccessible to the centers of the microparticles due to the hard part of the potential $\phi_{12}(r)$. The rippled area represents the structuration of the density $\rho(\vs{r})$ due to the presence of the macrospheres.}
\end{figure}

Consider an asymmetric binary mixture (we label with 1 the larger component and with 2 the other one) and let the interaction potential between two particles of the larger species $\phi_{11}(r)$ and the interspecies potential $\phi_{12}(r)$ have a square-well or square shoulder form, that is 
\begin{equation}
	\phi_{ij}(r) = \left\{
		\begin{array}{rl}
			\infty			& \text{if }	r 							<	\sigma_{ij} 							\\
			- \epsilon_{ij}	& \text{if }	\sigma_{ij}					\le	r	<		\lambda_{ij}\,\sigma_{ij} 	\\
			0				& \text{if }	\lambda_{ij}\,\sigma_{ij} 	\le	r										\\
		\end{array}
		\label{eq:SquareWellPotential}
	\right. ,
\end{equation}

where $\sigma_{ij}$ denotes the range of the hard core interaction, $\epsilon_{ij}$ is a positive (negative) energy measuring the depth (height) of the well (shoulder) and $\lambda_{ij}$ determines the width of the well (shoulder) with respect to the range of the hard core. We fix $\epsilon_{11} = 0$ and leave the interaction potential of the smaller species $\phi_{22}(r)$ unspecified. We now wish to calculate the effective force between two particles of type 1 immersed in a sea of particles of type 2.\\
The problem in the particular case of $\epsilon_{12} = 0$ (the hard sphere limit) has already been studied in detail \cite{attard1989spherically, dickman1997entropic} in the past. If two spheres of type 1 are fixed in $\vs{0}$ (the origin) and $\vs{R}$ respectively, the value $f$ of the radial component of the effective force between each other can be written as (\onlinecite{attard1989spherically}): 
\begin{equation}
	f(R) = - \frac{2 \pi}{\beta} \sigma_{12}^2 \int_0^{\pi} \rho (\vs{S}_{\sigma_{12}}; R) \cos{\theta} \sin{\theta} d\theta\ ,
	\label{eq:AttardForce}
\end{equation}

where $\beta$ is the inverse temperature $1/k_B T$, $\vs{S}_{\sigma_{12}}$ is a vector of length $\sigma_{12}$ whose tail is in $\vs{0}$ and $\theta$ is such that $\vs{S}_{\sigma_{12}} \cdot \vs{R} = R\, \sigma_{12} \cos \theta$ (the meaning of the parameters that appear in  \ref{eq:AttardForce} is further elucidated in  \ref{sec:ForceDerivationDetails} and \ref{fig:Parameters}).
Thus the force depends on the \textit{density} $\rho(\vs{r})$ of small particles around the pair of macrospheres. 

In  \ref{sec:ForceDerivationDetails} a derivation similar to Attard's is used to obtain the expression of the effective force between two macrospheres when the interaction between the species 1 and 2 is of the kind described by  \ref{eq:SquareWellPotential}:
\begin{align}
	f(R) & = - \frac{2 \pi}{\beta} \left(\sigma_{12}^2 \int_0^{\pi} \rho (\vs{S}_{\sigma_{12}}; R) \cos{\theta} \sin{\theta} d\theta\ + \right. \nonumber \\
&	\left. (1 - e^{\beta \epsilon_{12}}) \lambda^2 \sigma_{12}^2 \int_0^{\pi} \rho(\vs{S}_{\lambda \sigma_{12}}; R) \cos{\theta} \sin{\theta} d\theta \right) ,
	\label{eq:FinalForce}
\end{align}
where the shorthand notation $\lambda\, \sigma_{12} = \lambda_{12}\,\sigma_{12}$ has been used, $\vs{S}_{\lambda \sigma_{12}}$ is defined in a fashion similar to $\vs{S}_{\sigma_{12}}$ and a representation of all the parameters is given in \ref{fig:Parameters}.

Our expression differs from the hard sphere case ( \ref{eq:AttardForce}) because of the presence of a second term. This is due to the introduction of the square well (shoulder) of finite depth (height) $\abs{\epsilon_{12}}$ and has the same form of the previous term but a weight $1 - e^{\beta \epsilon_{12}}$, which is negative for square well potentials. This dependence of the force on the temperature is a by-product of the non-zero value of $\epsilon_{12}$, which introduces a natural energy scale that is absent in the athermal hard sphere system. Thus, the force is not entirely entropic as in  \ref{eq:AttardForce} but has also an energetic origin. 
The expression of the force in  \ref{eq:FinalForce} correctly reduces (as it should) to Attard's formula in the $\epsilon_{12} = 0$ (hard sphere) limit and to the same expression and a larger $\tilde{\sigma} = \lambda \sigma_{12}$ in the limit $\epsilon_{12} \to - \infty$ (infinitely high shoulder). We emphasize the fact that  \ref{eq:FinalForce} yields the \textit{exact} force if the correct form of the density is known regardless the interaction potential between the small particles. In this paper we shall restrict ourselves to the two cases in which the small particles are either non-interacting ($\phi_{22}(r) = 0$) or hard spheres (so that $\phi_{22}(r)$ will take the form of  \ref{eq:SquareWellPotential} with $\epsilon_{22} = 0$).

\section{\label{sec:DensityDetermination}Determination of the density and force}
In order to determine the effective force the density $\rho(\vs{r})$ was obtained using three different methods: the dilute gas approximation, integral equations together with the superposition approximation and Metropolis Monte Carlo simulations.

\subsection{\label{sec:DLapprox}Dilute limit (DL) approximation}
\begin{figure}
	\includegraphics[width = 0.6\textwidth]{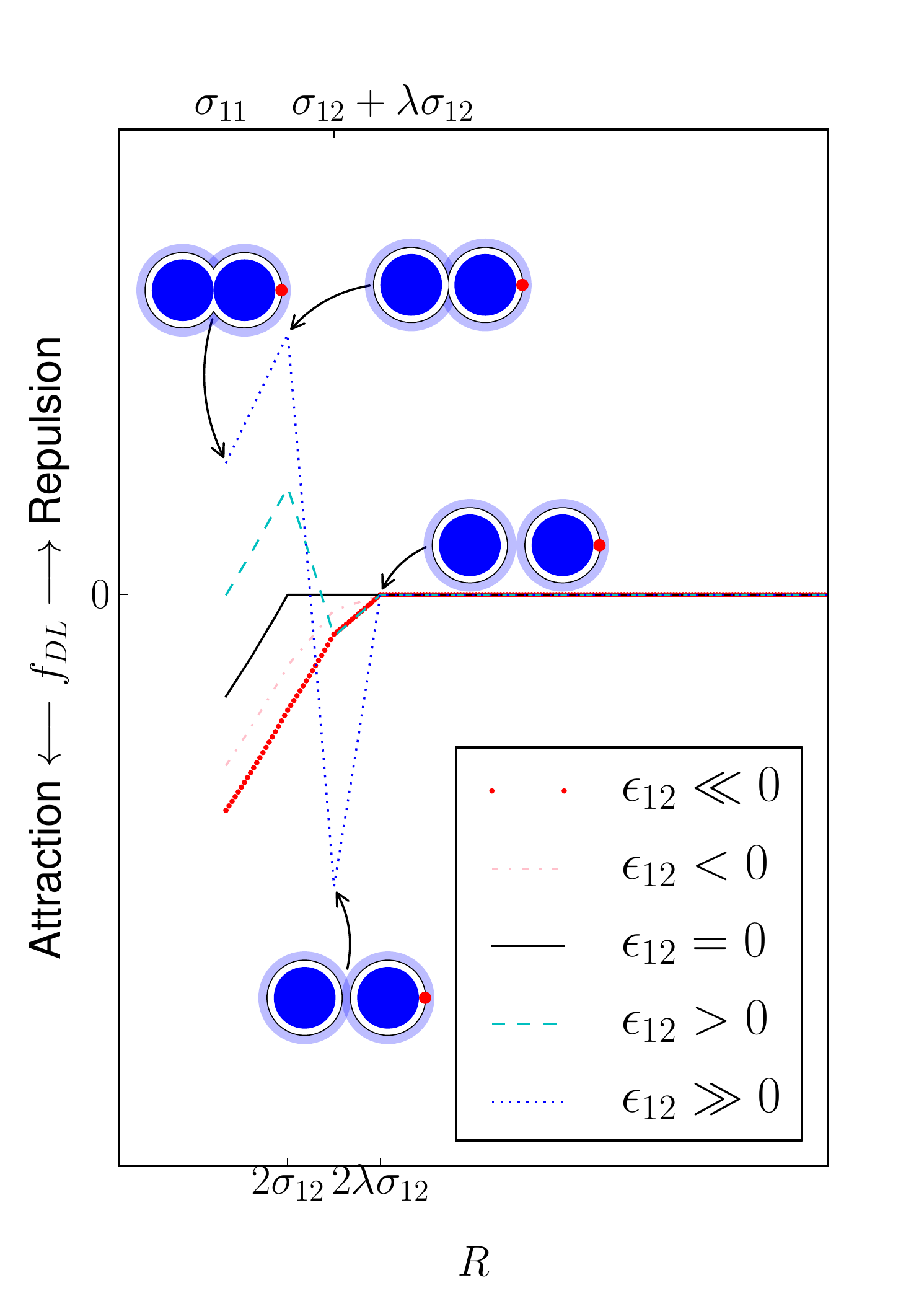}
	\caption{\label{fig:DLforce}Qualitative force profiles for various values of $\epsilon_{12}$ in the dilute limit. This qualitative behavior is observed in a quite broad region in the space of the geometrical parameters. The configurations corresponding to $R = \sigma_{11}, 2 \sigma_{12}, \sigma_{12} + \lambda \sigma_{12}, 2 \lambda \sigma_{12}$ are shown. The smaller disks represent the hard spheres, the white coronas the volume forbidden to the centers of the microspheres, and the bigger coronas are the square well/shoulder. The size of a microsphere is that of the red disks. The geometrical parameters used in the drawing are $\sigma_{11} = 5$, $\sigma_{12} = 3$, $\lambda \sigma_{12} = 3.75$.}
\end{figure}

If species 2 is dilute we can approximate the density around a pair of macrospheres with that of an ideal gas: 
\begin{equation}
	\rho_{\rm DL}(\vs{r}; R) = \rho e^{-\beta V(\vs{r}; R)},
	\label{eq:DLapprox}
\end{equation}
where $V(\vs{r}; R)$ is the potential energy of a particle of type 2 centered in $\vs{r}$ due to the presence of the fixed pair of type 1 (see also  \ref{eq:ExternalPotentialDetail}). \\
This is equivalent to the Asakura-Oosawa approximation in the case $\epsilon_{12} = 0$. We expect such approximation to be less accurate in more coupled regimes but still able to give qualitative information about the force profile.
In App.~\ref{sec:nastyDLforce} we make use of  \ref{eq:DLapprox} and obtain an \textit{analytic} expression of the effective force. For a wide choice of the parameters $\epsilon_{12}, \sigma_{12}, \lambda \sigma_{12}$ it has features like those appearing in \ref{fig:DLforce}.
The profile obtained when $\epsilon_{12} = 0$ is exactly the same short range depletion attraction given by the Asakura-Oosawa approximation \cite{attard1989spherically}. The effect of \textit{negative} values of $\epsilon_{12}$ is that of increasing both the range and the strength of the effective force, approaching the $\epsilon_{12} = - \infty$ solution mentioned above. In this case the qualitative features are essentially the same as in the $\epsilon_{12} = 0$ case.
\textit{Positive} values of $\epsilon_{12}$ determine a completely different behavior: increasingly large values determine the onset of a repulsion at short distances (maximum repulsion occurring at $R = 2 \sigma_{12}$) and of a strong attractive force at intermediate distances (with maximum attraction found at $R = \sigma_{12} + \lambda \sigma_{12}$).

\begin{figure}[!ht]
	\centering
	\subfigure[\label{fig:FarHardSpheres}]{\includegraphics[width = 0.48\textwidth ]{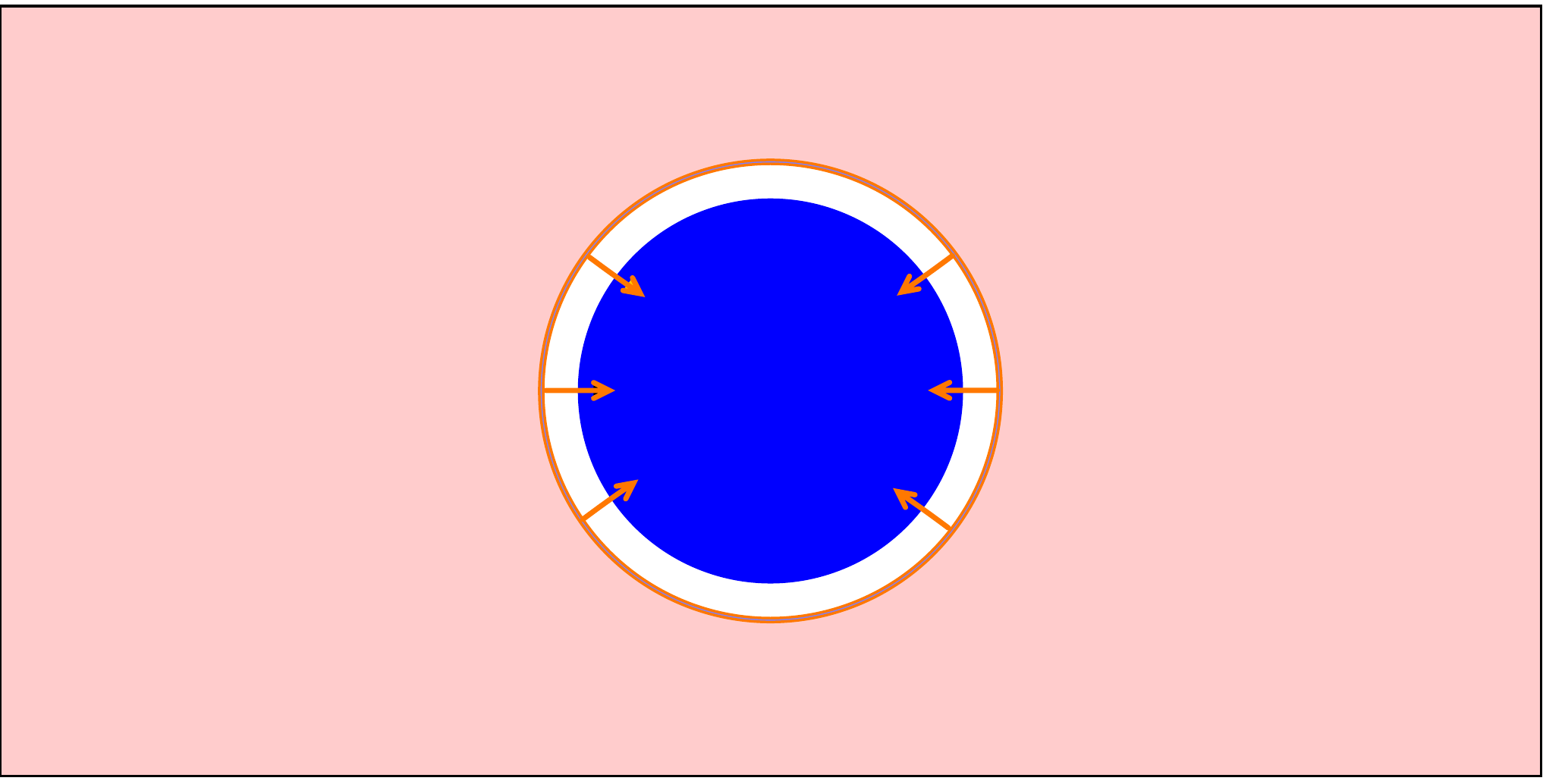}}
	\subfigure[\label{fig:NearHardSpheres}]{\includegraphics[width = 0.48\textwidth]{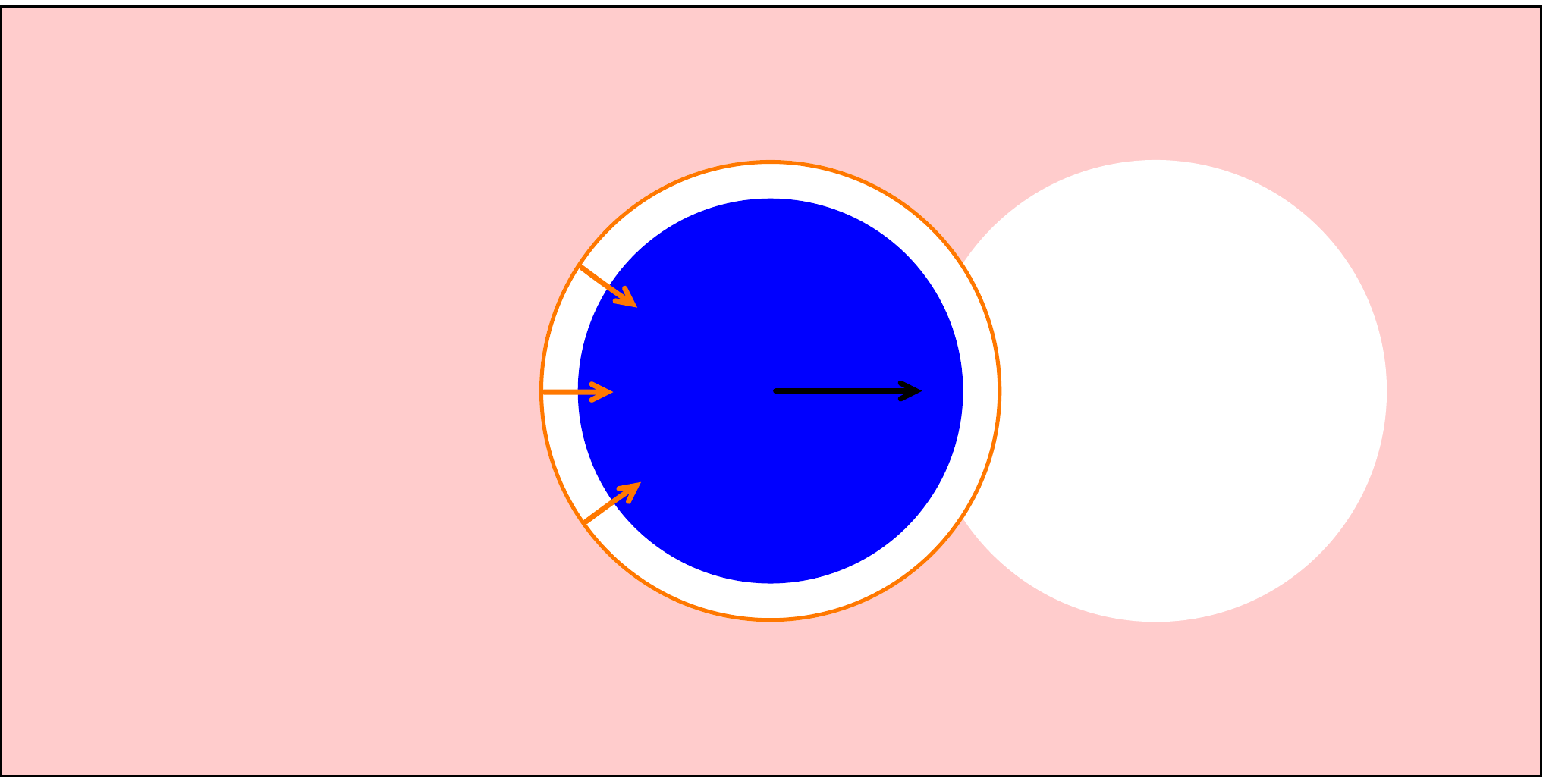}}
	\subfigure[\label{fig:FarSquareShoulder}]{\includegraphics[width = 0.48\textwidth ]{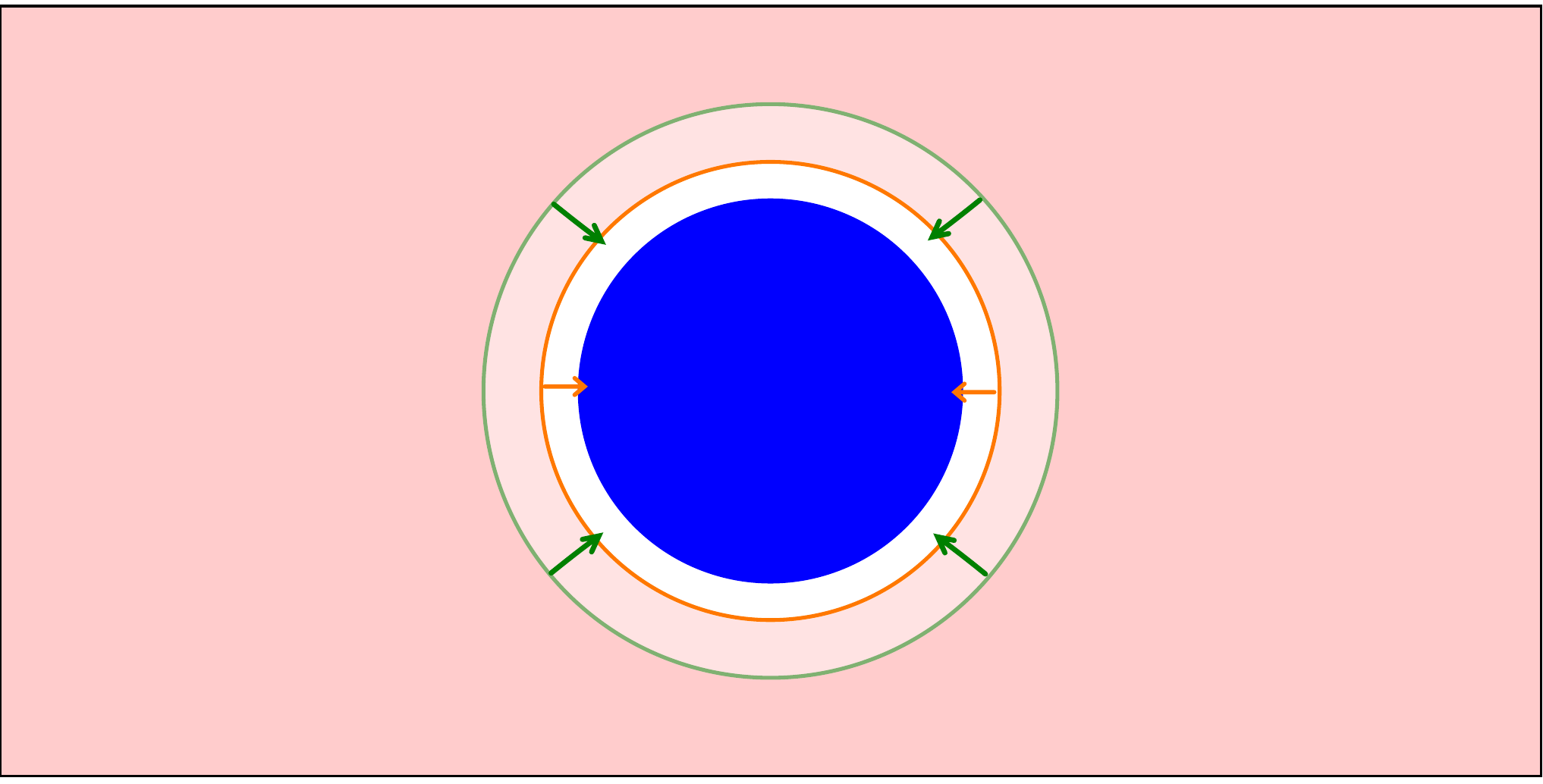}}
	\subfigure[\label{fig:NearSquareShoulder}]{\includegraphics[width = 0.48\textwidth]{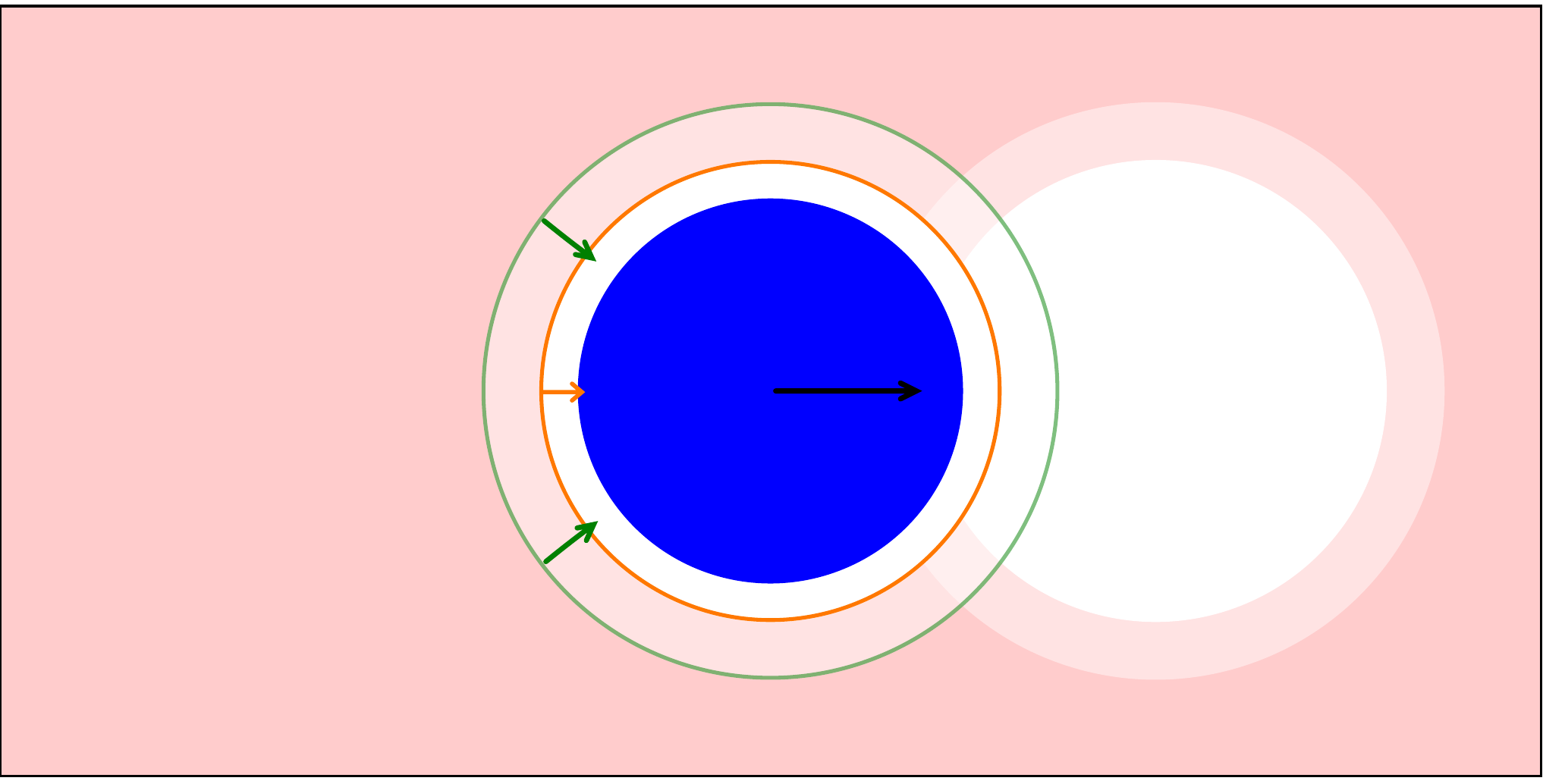}}
	\caption{\ref{fig:FarHardSpheres}, \ref{fig:NearHardSpheres}: Plot of the density of small particles around a big sphere (the blue disk) at large $R$ (isolated particle case) and when another sphere (not shown) is present at small $R$ in the case $\epsilon_{12} = 0$ and in the dilute limit. The anisotropy of the density determines the onset of an attractive force.\\
	 \ref{fig:FarSquareShoulder}, \ref{fig:NearSquareShoulder}: Density of small particles around a macrosphere at large $R$ and small $R$ in the case $\epsilon_{12} < 0$. The anisotropy of the density determines the onset of an attractive force as in  \ref{fig:NearHardSpheres}, this time due to the contribution of the density both at the surfaces of radius $\sigma_{12}$ (hard core) and $\lambda \sigma_{12}$ (outer rim of the shoulder). 
	\label{fig:HardSpheresSquareShoulderDLGraphicalExplanation}
	}
\end{figure}

The physical origin of such form of the effective interaction can be understood (at least qualitatively) by means of the argument below.\\
For sake of simplicity, let's consider the $\epsilon_{12} = 0$ case first. 
In this case one macroparticle and a microparticle \textit{repel} each other when they come into contact, i.e. when they are at a distance $r = \sigma_{12}$. It follows that if a macrosphere is inserted in a sea of small ones, it will experience on average a force depending on the density of small spheres at distance $r = \sigma_{12}$. If a macrosphere is isolated from others the density distribution of small ones will be spherically symmetric so it won't experience any net force. If two macrospheres come close together, however, the hemispheres at $r = \sigma_{12}$ facing each other become depleted of microspheres: the ``push'' from the two external hemispheres is not balanced anymore and the result is an effective attraction. The situation is depicted in \ref{fig:HardSpheresSquareShoulderDLGraphicalExplanation}.

Now consider the case when a square well or shoulder is also present. We have to take into account that a macroparticle and a microparticle experience an \textit{attraction or a repulsion depending on the sign of $\epsilon_{12}$} when their separation is $\lambda \sigma_{12}$. Thus, the effect of a well is to make a macrosphere being \textit{attracted} by the density of small spheres localized at a distance $r = \lambda \sigma_{12}$, while a shoulder will make the macrosphere be pushed away from it. 
Taking this into account we can understand the effect of $\epsilon_{12}$ on $f_{\rm DL}(R)$. \\
It's easy to realize (see also \ref{fig:HardSpheresSquareShoulderDLGraphicalExplanation}) that the case $\epsilon_{12} < 0$ (shoulder) is similar to that where $\epsilon_{12} = 0$, with the novel contribution due to the density localized at the outer rim of the shoulder and a weaker contribution from the hard wall due to lower density of small particles inside the shoulder.

\begin{figure}[!ht]
	\centering
	\subfigure[\label{fig:FarSquareWell}]{\includegraphics[width = 0.48\textwidth]{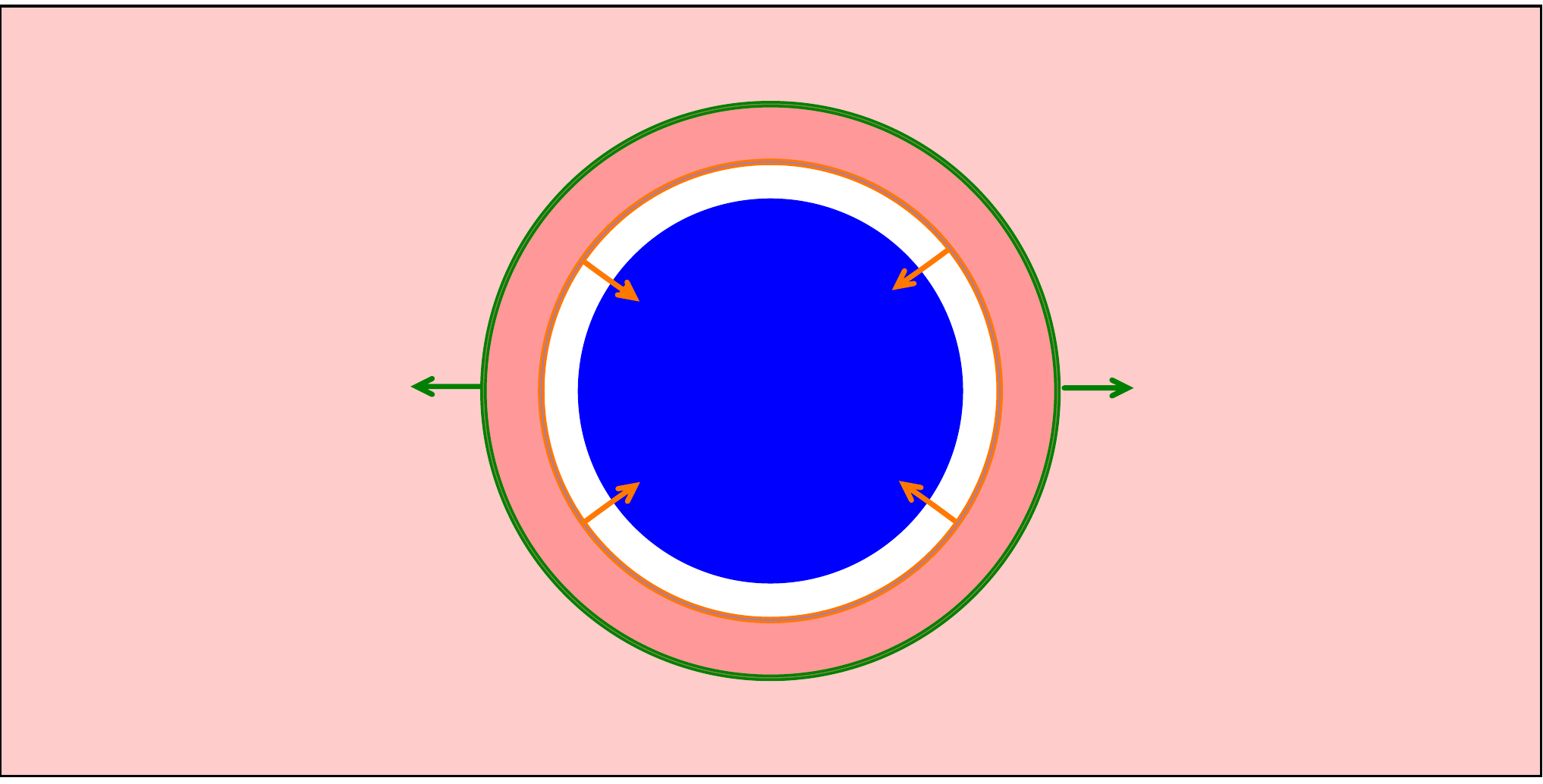}}
	\subfigure[\label{fig:SemifarSquareWell}]{\includegraphics[width = 0.48\textwidth]{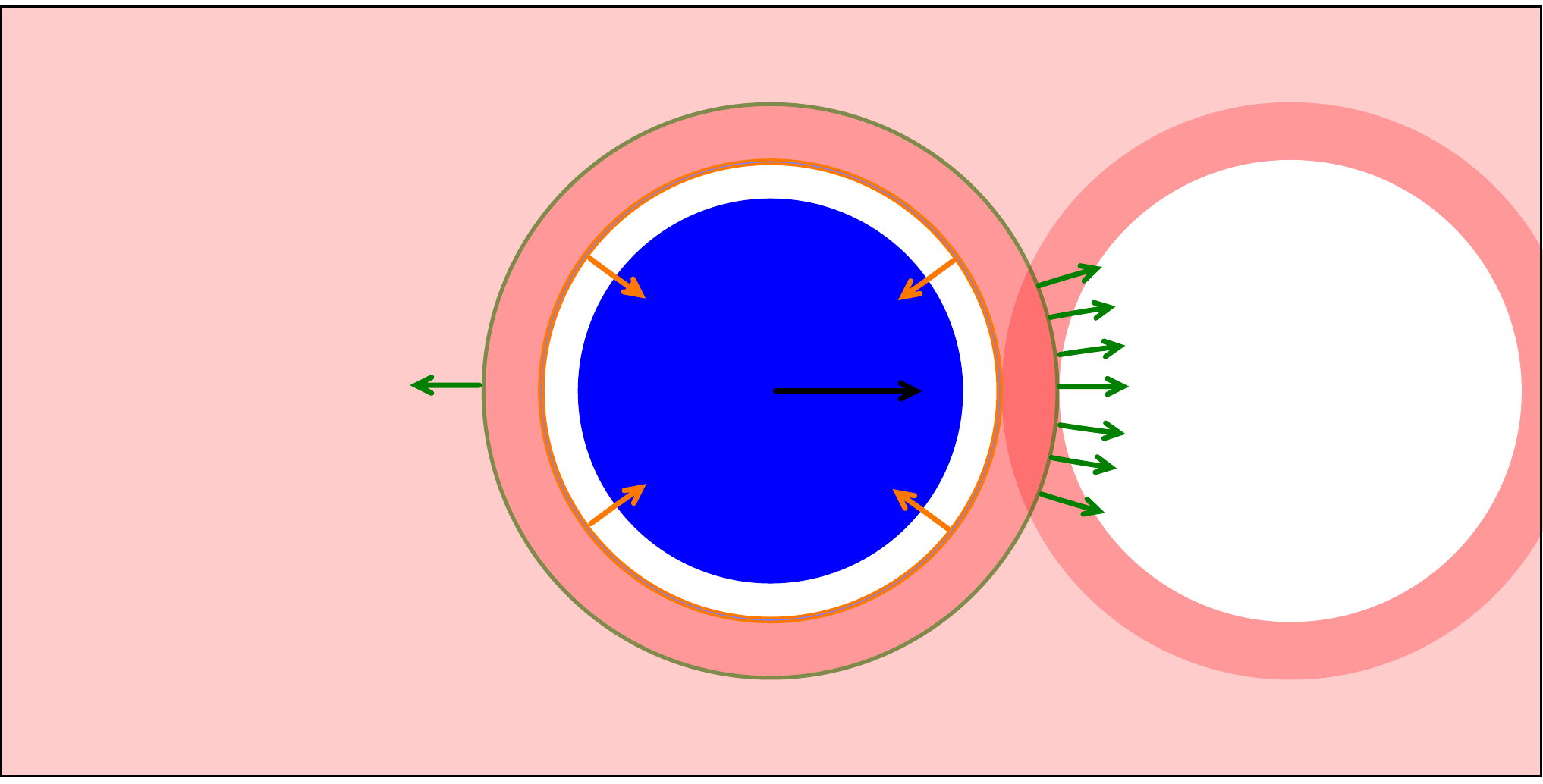}}
	\subfigure[\label{fig:MedSquareWell}]{\includegraphics[width = 0.48\textwidth]{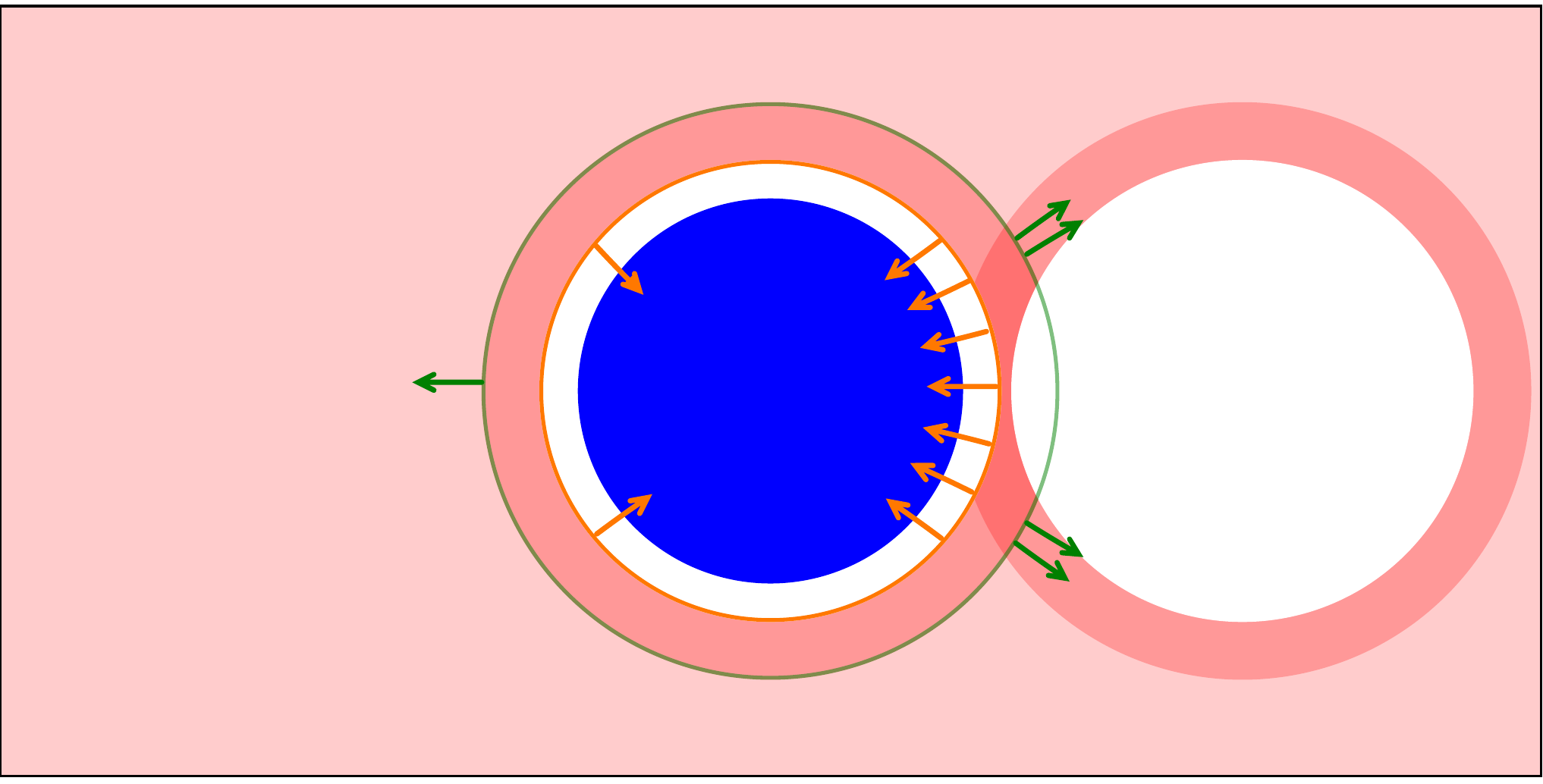}}
	\subfigure[\label{fig:NearSquareWell}]{\includegraphics[width = 0.48\textwidth]{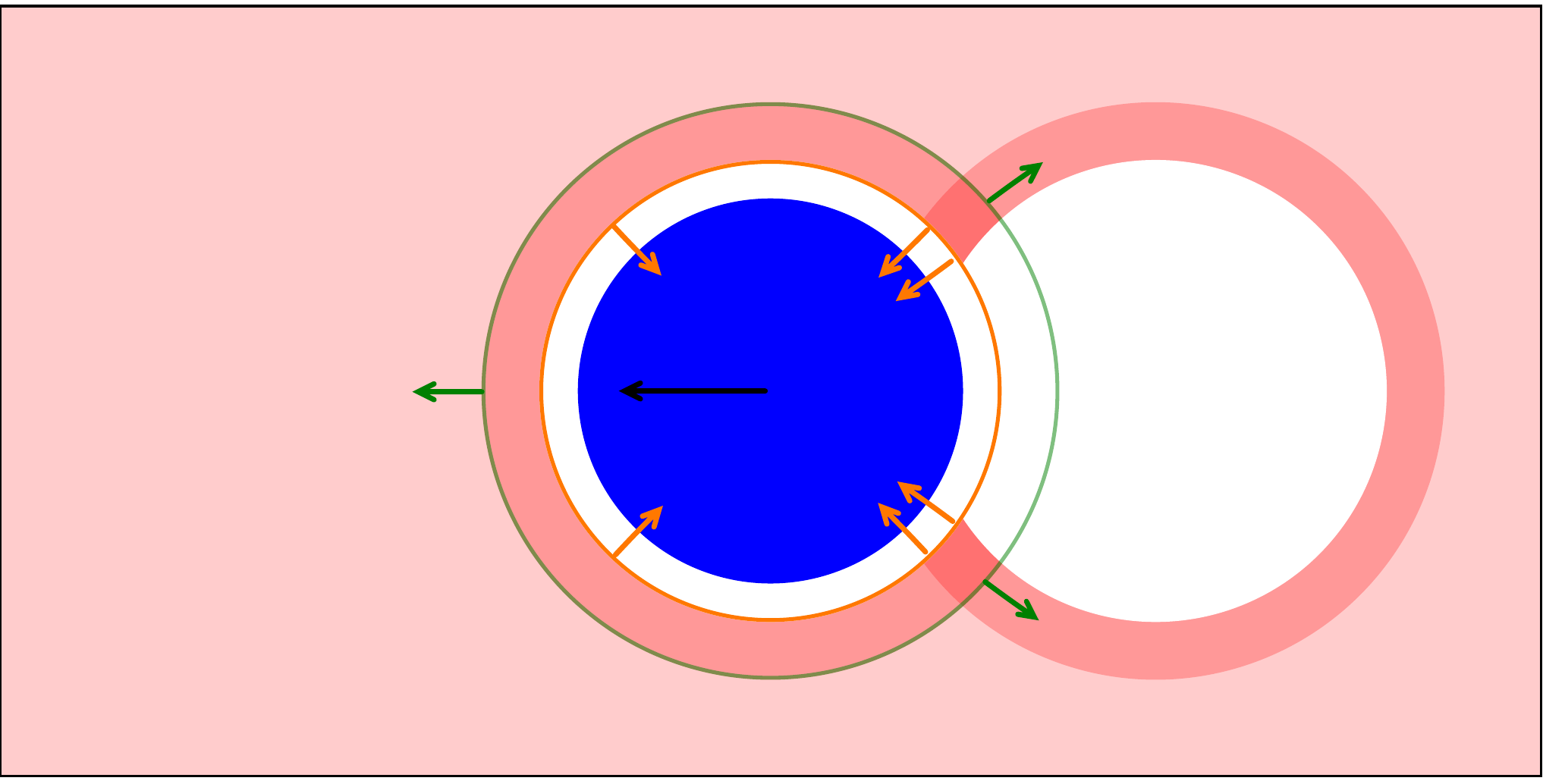}}
	\caption{Density of small particles around a macrosphere for decreasing $R$ in the case $\epsilon_{12} > 0$. The effective interaction is the result of the interplay between the ``push'' due to the density at the surface of radius $\sigma_{12}$ (much stronger due to the higher density inside the well) and the ``pull'' at $\lambda \sigma_{12}$ (outer rim of the well). 
	\label{fig:SquareWellDLGraphicalExplanation}
	}
\end{figure}

The situation is more complicated if $\epsilon_{12} > 0$ (well) and is schematized in \ref{fig:SquareWellDLGraphicalExplanation}.
If $\sigma_{12} + \lambda \sigma_{12} < R < 2 \lambda \sigma_{12}$ part of the outer rim of the well is inside the well of the neighboring big particle and there is a strong ``pull'' due to the high density of small particles here (green arrows in  \ref{fig:SemifarSquareWell}). The result is an effective attraction. At lower $R$ such attraction is counterbalanced by a strong ``push'' due to the fact that the density at the hard wall is higher where the two cores face each other (see blue arrows in  \ref{fig:MedSquareWell}). At the same time the attractive ``pull'' at the outer rim of the well (green arrows on the right in  \ref{fig:MedSquareWell}) is weaker because part of it is in the region not accessible to the small spheres ($\rho = 0$). For these reasons at small $R$ and big enough $\epsilon_{12}$ repulsion dominates over attraction ( \ref{fig:NearSquareWell}). 

\subsection*{Qualitative dependence of the phase diagram from $\epsilon_{12}$}
We studied the effect of the presence of the smaller species on the phase diagram of the larger component, to see if the stability of the system can be tuned by means of varying the parameter $\epsilon_{12}$. Effects of the mutual attraction on the stability of this class of systems has previously been reported in the literature \cite{stradner2007new, dorsaz2008colloidal}. 
In the semi-grand canonical ensemble we can write for the thermodynamic potential $F(N, V, z_2)$
\begin{equation}
e^{-\beta F} = \frac{1}{N! \Lambda_1} {\rm Tr_1} e^{-\beta H_{\rm eff}},
\label{eq:Semigrandpartition}
\end{equation}
where $N$ is the number of large particles, $V$ is the volume of the system, $z_2$ is the fugacity of the smaller species \cite{dijkstra1999phase}, $\rm Tr_1$ denotes the integration operator over the coordinates of the particles of the larger species $\int \vs{dR_1 \ldots dR_N}$, $\Lambda_1 \equiv h/\sqrt{2 \pi m_1 / \beta}$ its thermal wavelength and $H_{\rm eff}$ is the effective potential. $H_{\rm eff}$ in turn can be written as
\begin{equation}
H_{\rm eff} = H_{11} + \Omega,
\label{eq:EffectivePotential}
\end{equation}
where $H_{11}$ is the sum of the pair interactions between the larger particles $\sum_{i \neq j}^N \phi_{11}(R)$ and $\Omega$ is the grand potential of the smaller species in a fixed configuration of the large particles.
It can be shown that $\Omega$ can be expressed as a sum of $n$-body terms \cite{likos2001effective, dijkstra1999phase}
\begin{equation}
\Omega = \sum_{n} \Omega_n
\label{eq:OmegaExpansion}
\end{equation}
 and it is straightforward (see  \ref{sec:VolumeTerms}) to prove that the 0- and 1-body terms are linear in the density of the effective component $\rho_1$ and therefore do not alter the phase behavior \cite{likos2001effective, dijkstra1999phase} in the case examined here.
In what follows the 2-body term is assumed to be given by 

\begin{equation}
\Omega_2 = \sum_{i \neq j}^N V_{\rm DL}(R_{ij}),
\label{eq:2BodyTerm}
\end{equation}

where

\begin{equation}
V_{\rm DL}(R) = - \int f_{\rm DL}(R) dR + c,
\label{eq:DLpotential}
\end{equation}
where $c$ is chosen to guarantee that $\lim_{r \to \infty} V_{\rm DL}(r) = 0$ holds.

First order thermodynamic perturbation theory \cite{hansen2006theory} can thus be employed to explore the phase behavior of the effective component, in a way similar to that applied by Gast and coworkers in the AO case ($\epsilon_{12} = 0$) \cite{gast1983polymer}.
We thus write the free energy per particle of the effective component $F^{\rm eff}/N$ with
\begin{equation}
	\beta F^{\rm eff}/N = \beta F_{\rm HS}/N + \frac{\beta \rho_1}{2} \int V_{\rm DL}(r)\, g_{\rm HS}(r)\, 4 \pi r^2 dr,
	\label{eq:FirstOrderFreeEnergy}
\end{equation}

where $F_{\rm HS}/N$ and $g_{\rm HS}$ are respectively the free energy per particle and the pair distribution function associated to the hard sphere reference system. The former is approximated here by means of the Carnahan-Starling \cite{hansen2006theory} expression and the Verlet-Weis \cite{verlet1972equilibrium} description is used for the latter.
A check of the validity of the first-order approach can be performed by verifying that the Barker-Henderson second-order correction to $F^{\rm eff}/N$ \cite{gast1983polymer, barker1967perturbation} is reasonably smaller than the second term on the RHS of  \ref{eq:FirstOrderFreeEnergy}.
The chemical potential $\mu$ and the pressure $p$ can be expressed with 

\begin{equation}
\beta \mu = \frac{\partial}{\partial \rho_1} (\rho_1 \beta F^{\rm eff}/N),
	\label{eq:ChemicalPotential}
\end{equation}

\begin{equation}
\beta p = \rho_1 \beta \mu - \rho_1 \beta F^{\rm eff}/N.
	\label{eq:Pressure}
\end{equation}

Coexistence lines are found by plotting the value of the volume fraction $\eta_1$ of the bigger spheres at which the parametric curve in the $(p, \mu)$ plane self-intersects for various values of the volume fraction $\eta$ of the smaller particles.

\subsection{\label{sec:IE} Integral equations and superposition approximation (PY + S  and HNC + S) methods}
Another approximate way to determine the density is using integral equation theory for a fluid mixture in order to get the unlike (big-small) pair distribution $g_{12}(r)$. This in turn can be plugged into the superposition approximation for the density profile around the pair of macrospheres
\begin{equation}
	\rho_{\rm S}(\vs{r}; R) = g_{12}(r) g_{12}(\abs{\vs{r} - \vs{R}}) \rho.
	\label{eq:SuperpositionApproximation}
\end{equation}
Note how the superposition approximation amounts to saying that the density around the pair is equal to the product of the densities that the particles would have around themselves if they were isolated.

The pair distribution function $g_{12}$ in  \ref{eq:SuperpositionApproximation} can be determined solving the two-component Ornstein-Zernike equation
\begin{equation}
	h_{ij}(r) = c_{ij}(r) + \sum_{k = 1}^2 \int \rho_k\, c_{ik}(\abs{\vs{r} - \vs{r'}})\, h_{kj}(r')\, \vs{dr'},
	\label{eq:OZ}
\end{equation}
and the closure equation
\begin{equation}
	g_{ij}(r) = e^{-\beta \phi_{ij}(r) + h_{ij}(r) - c_{ij}(r) - E_{ij}(r)},
	\label{eq:Closure}
\end{equation}
where $i,\ j,\ k \in \{1, 2\}$,  $h_{ij}, c_{ij}, E_{ij}(r)$ are respectively the indirect correlations, direct correlations and bridge functions \cite{caccamo1996integral}, $\rho_k$ is the density of the species $k$ and $\sum_k \rho_k = \rho$.
In order to solve the integral equations we implemented a version of Gillan algorithm \cite{gillan1979new}. Special care was taken to treat discontinuous potentials of the form of  \ref{eq:SquareWellPotential}. 

The well-known Percus-Yevick closure (PY)
\begin{equation}
	E_{ij}(r) = \ln [1 + h_{ij}(r) - c_{ij}(r)] - h_{ij}(r) + c_{ij}(r)
	\label{eq:PYBridge}
\end{equation}
and the hypernetted-chain closure (HNC)
\begin{equation}
	E_{ij}(r) = 0
	\label{eq:HNCBridge}
\end{equation}
were chosen.

\subsection{\label{sec:MC}Monte Carlo (MC) method}
The exact density can be sampled using Metropolis Monte Carlo simulations. The force can then be obtained with the method used in the two-sphere studies in (\onlinecite{dickman1997entropic}) (a possibly more efficient alternative method is that described in (\onlinecite{wu1999monte})). We briefly summarize such method here. Two macrospheres of diameter $\sigma_{11}$ are inserted at fixed positions $(0, 0, 0)$ and $(R, 0, 0)$ in a cell whose dimensions are $H$ along the $x$ direction and $L$ in the $y$ and $z$ directions. 
The same cell contains $N$ smaller particles which interact with each other with a potential of the form of  \ref{eq:SquareWellPotential} such that $\epsilon_{22} = 0$. $H$ and $L$ are chosen so that the density profile is flat away from the pair of macrospheres (i.e. in what we can consider to be \textit{the bulk})  and is equal to $\rho$. 
To keep $\rho$ constant for each value of $R$, the number $N$ is tuned to compensate the variation in the volume accessible to the small spheres.
Periodic boundary conditions and the NVT ensemble are used.\\
At the beginning of the simulation microparticles are placed randomly. At each MC step a microsphere is selected at random and a random displacement is attempted. The displacement is accepted or rejected according to the Metropolis scheme, the displacement is tuned to reach an acceptance ratio of 0.25 and cell lists are used in order to improve the efficiency of the algorithm \cite{frenkel2002understanding}. Configuration samples are taken after the mean square displacement of the microspheres equals $\sigma_{22}^2$ or after a number of moves sufficient to decorrelate the total energy per particle (in the SW and SS case). 
The value of the integrals that appear in  \ref{eq:FinalForce} can be obtained using the relation
\begin{equation}
	2\pi v^2 \int_0^{\pi} \rho(\vs{S_v}; R)\, \sin \theta \cos \theta \, d\theta \approx \big\langle{ \frac{1}{dr} \sum_{v < v_i < v + dv} \cos \theta_i \big\rangle}, 
	\label{eq:IntegralFromAverageCosinesSum}
\end{equation}
where we sum all the $\cos \theta_i$ whose distance $v_i$ from the center of the macrosphere at position $(R, 0, 0)$ is in the interval $[v, v + dv]$ and the angle brackets denote an average over all the samples collected during a simulation.
To obtain a better estimate of the integrals in  \ref{eq:FinalForce} a third-order polynomial was fitted with the RHS of  \ref{eq:IntegralFromAverageCosinesSum} corresponding to different $v$'s and extrapolated the curves to $v = \sigma_{12}^+$ and $v = \lambda \sigma_{12}^+$. 
The errors of the estimates are evaluated on the basis of the errors of the fit parameters. Data points in the fits were weighted according to the statistical uncertainties of the measured values of the RHS of  \ref{eq:IntegralFromAverageCosinesSum}.\\

\subsection*{\label{sec:DensityPlots}Plots of the density}

It's useful to obtain plots of the density obtained with the different methods in order to highlight any differences between their predictions. Such plots can be obtained from MC simulations by dividing the simulation box in cells, counting the number of particles contained in each of them, and averaging on multiple configurations. This data can be projected in 2D afterwards exploiting the symmetries of the system (for example $\rho(\vs{r}; R) \equiv \rho(r, \theta, \phi; R) = \rho(r, \theta; R)$ if $\phi$ defines a rotation around the direction of $\vs{R}$).

Plots of the density associated to the integral equation and superposition method can be obtained using  \ref{eq:SuperpositionApproximation} by means of sampling it on a very fine mesh in real space. Down-sampling to the same mesh used with MC data produces results that can be compared to those obtained with the MC method.

\subsection{Numerical details}
\begin{figure}[!ht]
	\subfigure{\includegraphics[width=.48\textwidth]{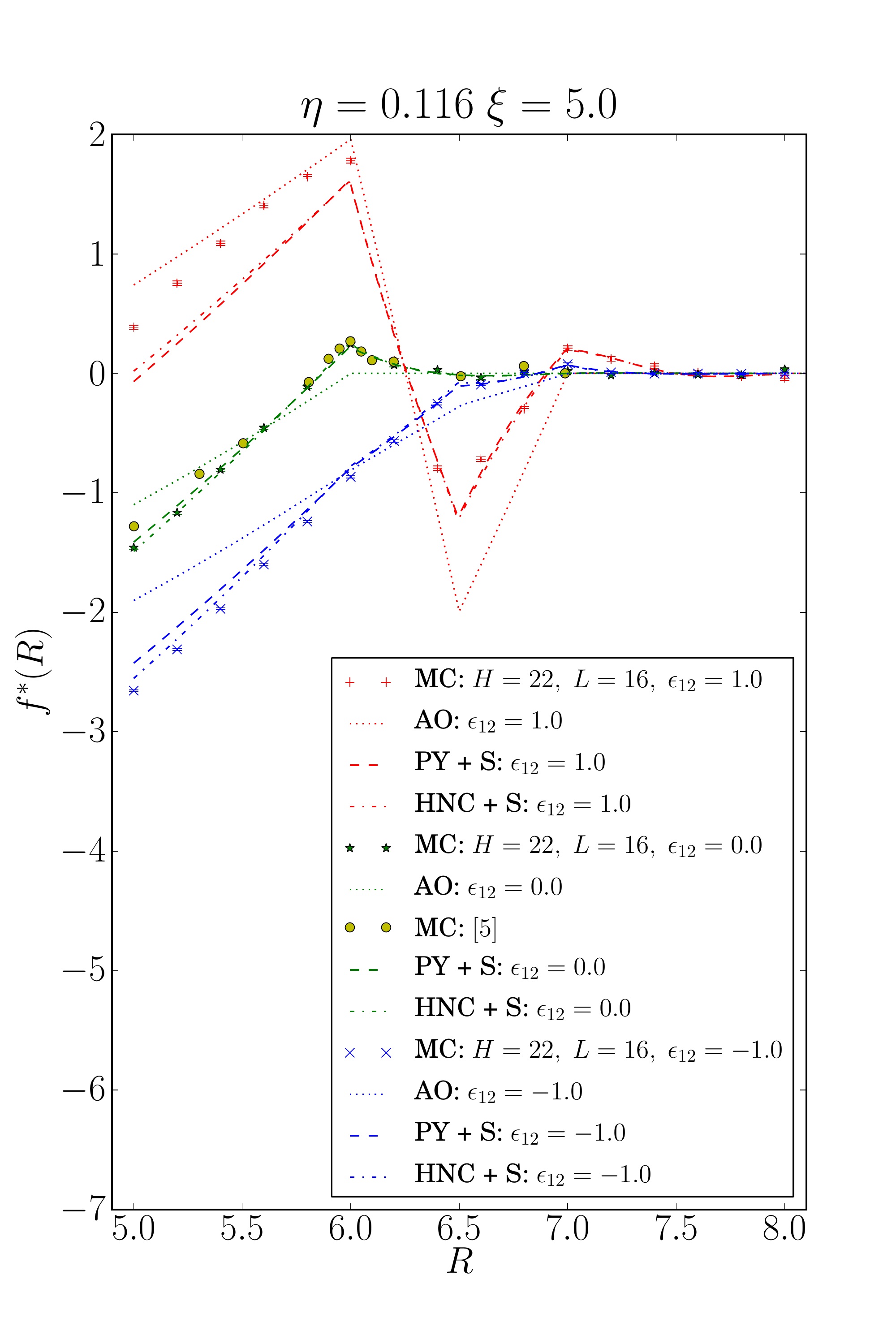}}
	\subfigure{\includegraphics[width=.48\textwidth]{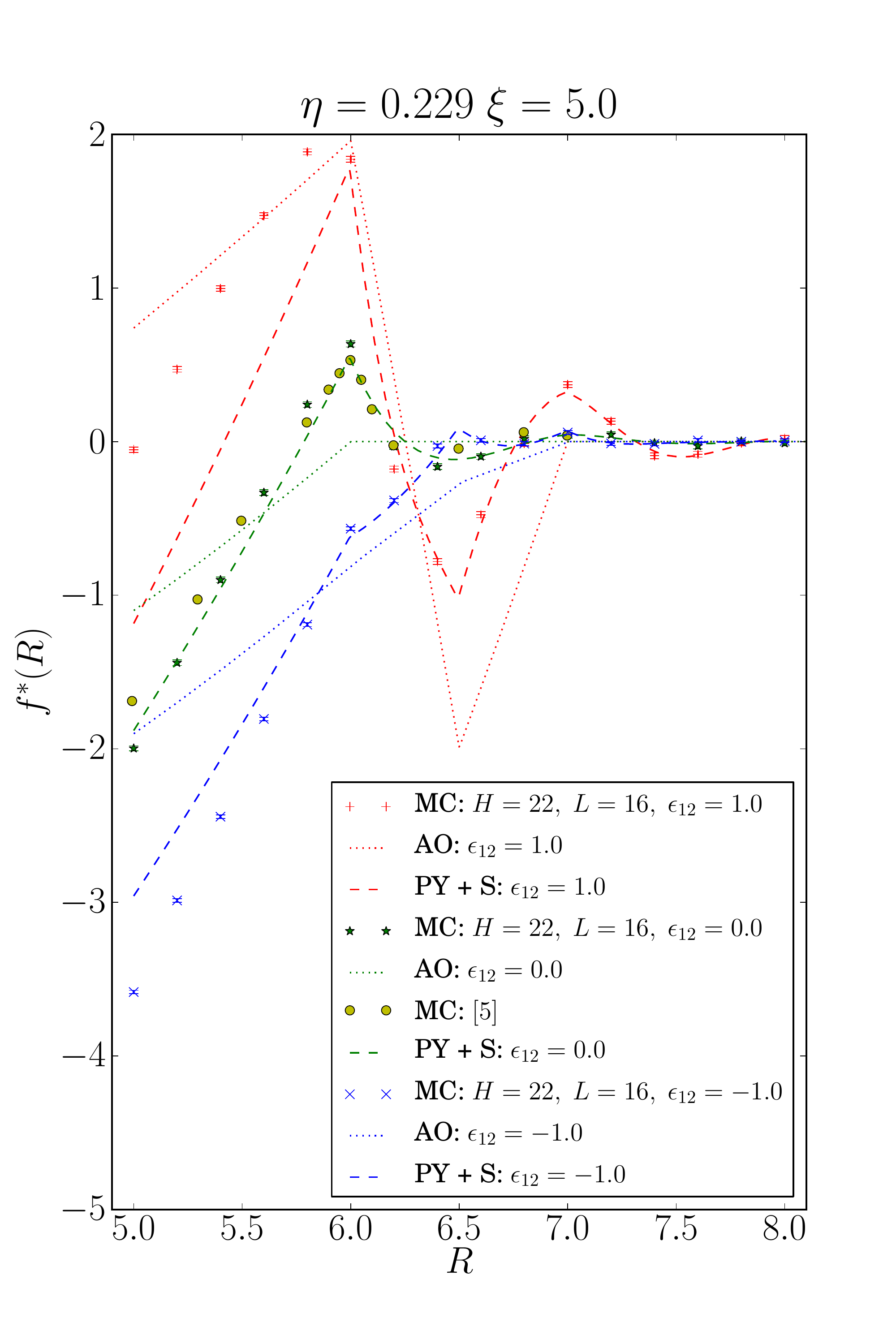}}
	\subfigure{\includegraphics[width=.48\textwidth]{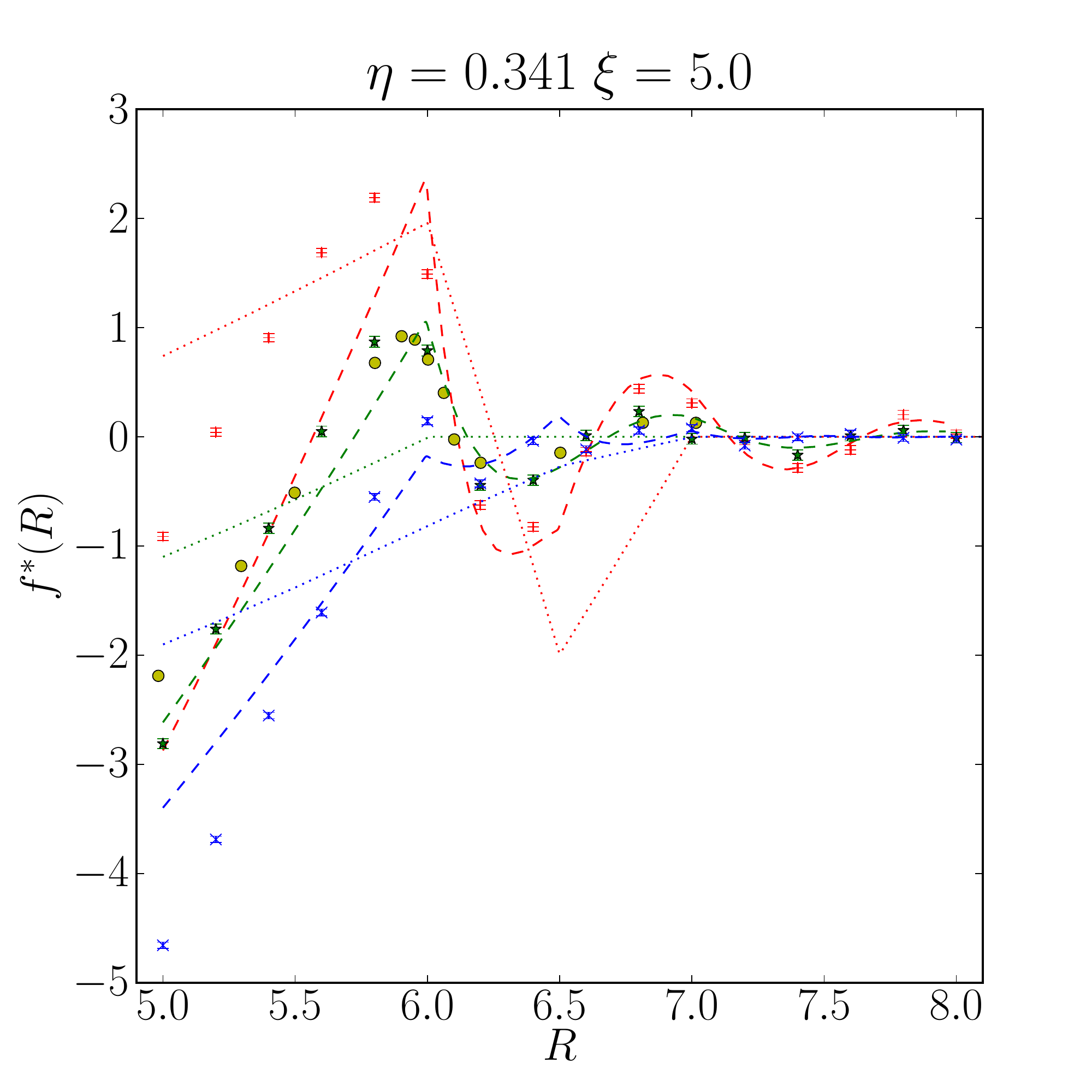}}
	\caption{Force profiles obtained with the various methods in the case $\xi = 5$, $\eta = 0.116,\ 0.229,\ 0.341$ for different values of $\epsilon_{12}$. In the case $\eta = 0.116$ the force profiles obtained with the HNC closure and the superposition approximation are also presented. The legend that holds for all the other force profiles in the article is that shown for the case $\eta = 0.229$. \label{fig:csi5}}
\end{figure}

\begin{figure}[!ht]
	\subfigure{\includegraphics[width=.48\textwidth]{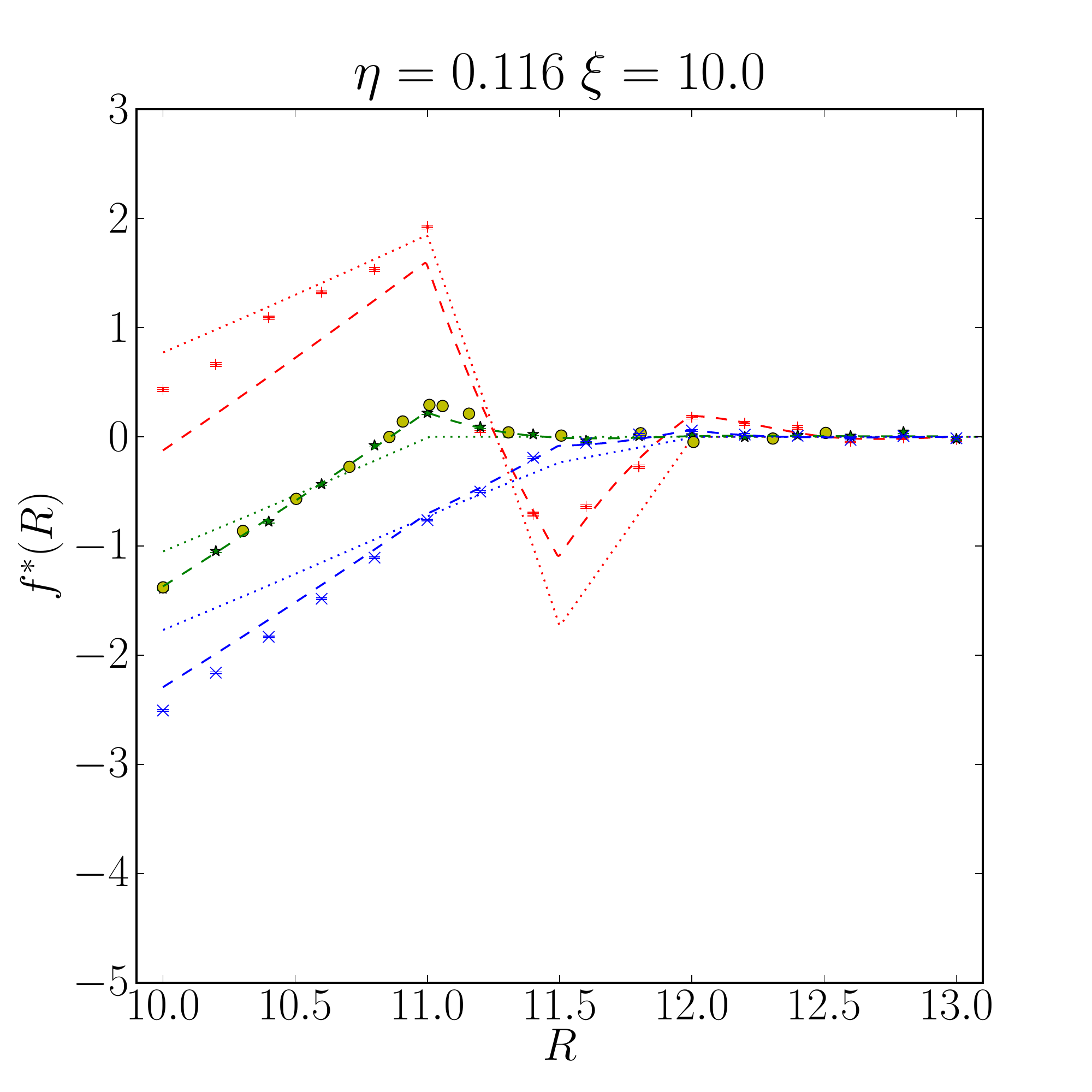}}
	\subfigure{\includegraphics[width=.48\textwidth]{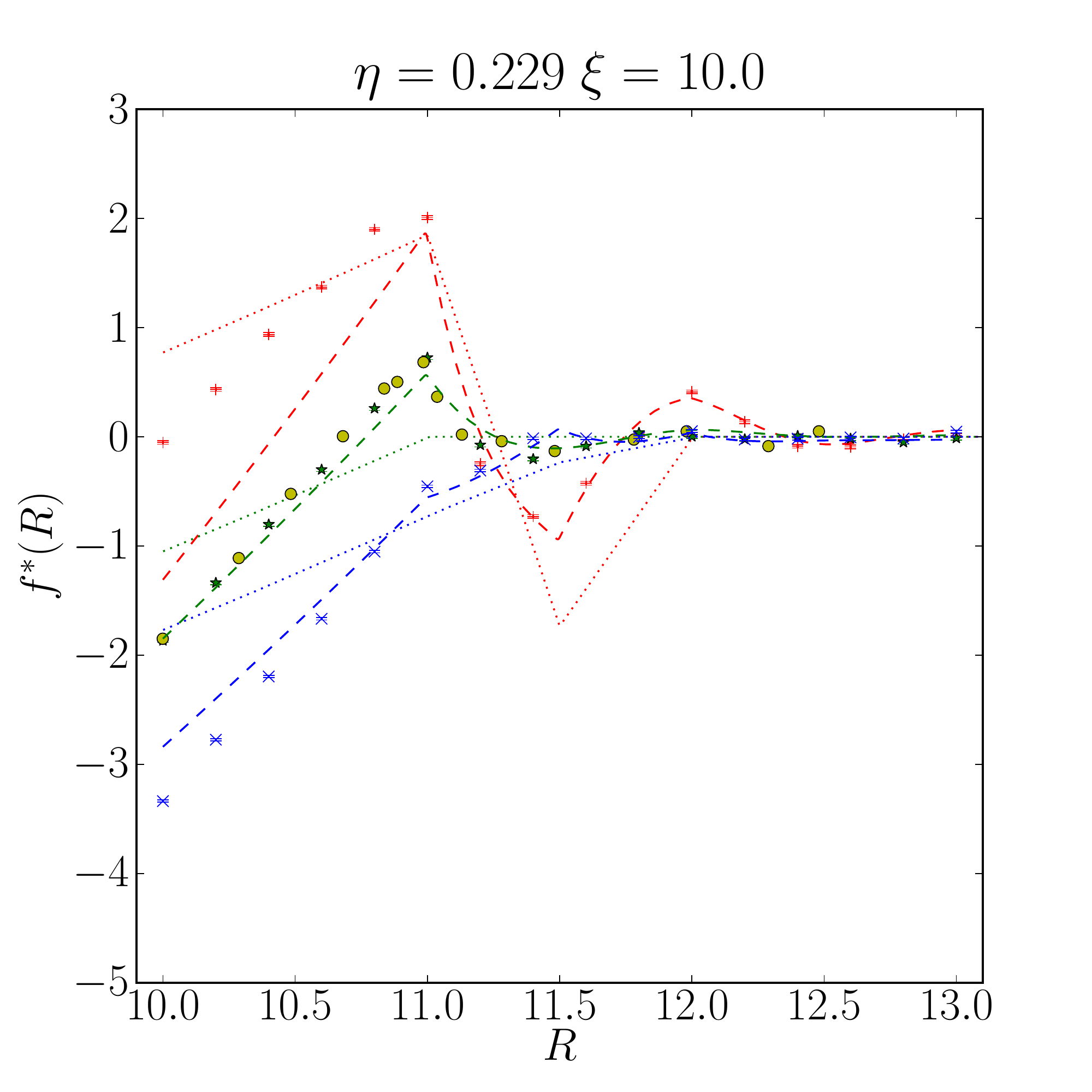}}
	\subfigure{\includegraphics[width=.48\textwidth]{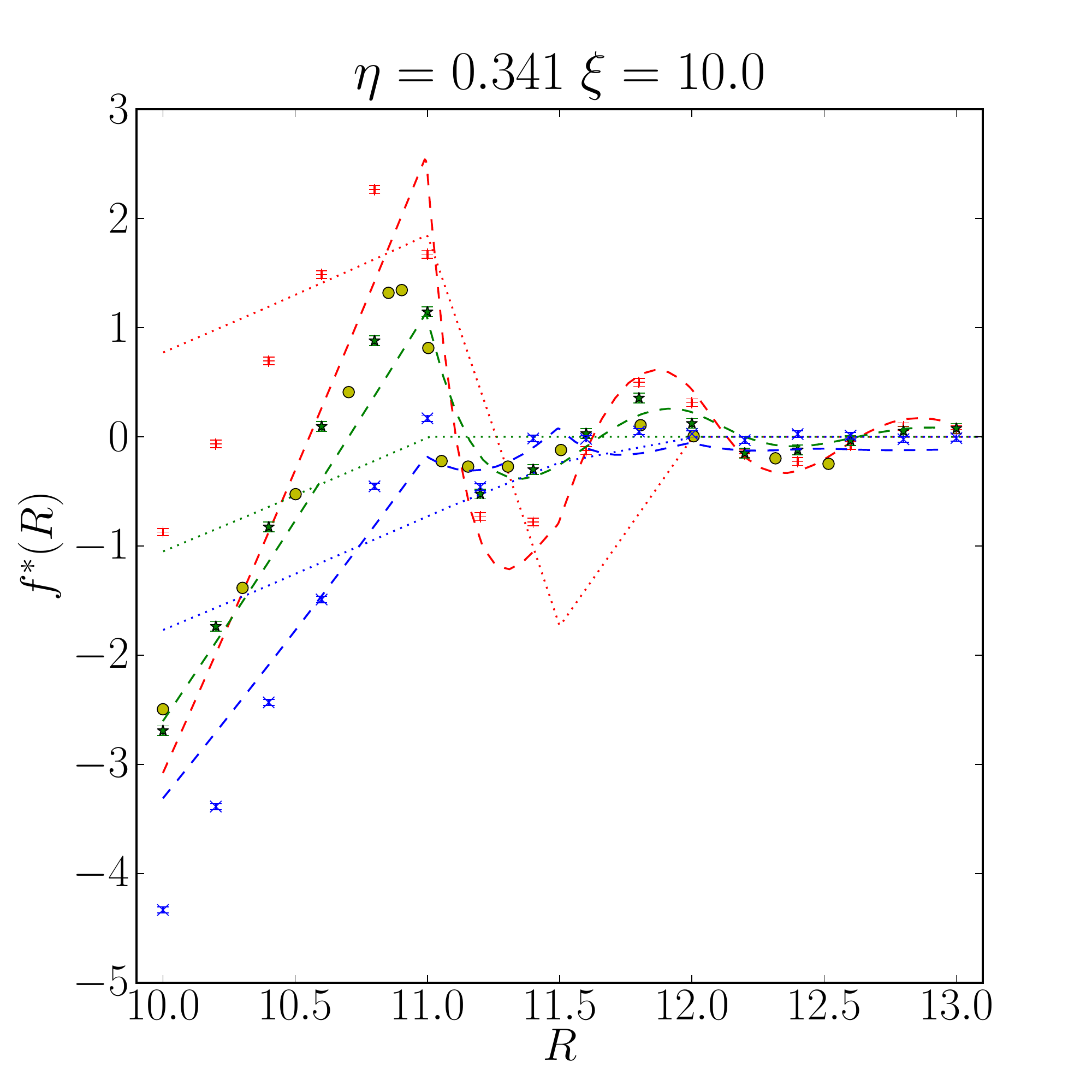}}
	\caption{Same as \ref{fig:csi5}, but using $\xi = 10$\label{fig:csi10}. MC data are obtained using a simulation box whose $H = 30$ and $L = 24$.}
\end{figure}

We used the same geometries analyzed by Dickman et al. in \onlinecite{dickman1997entropic}, i.e. two size ratios $\xi \equiv \sigma_{11}/\sigma_{22}$ = 5, 10, taking $\phi_{22}$ to be of the form of  \ref{eq:SquareWellPotential} with $\sigma_{22} = 1$ and $\epsilon_{22} = 1$ (that is, the smaller species is formed by unit hard spheres). The bulk packing fractions of the microspheres were also chosen in order to match those in (\onlinecite{dickman1997entropic}): $\eta = \pi \rho \sigma_{22}^3/6 = 0.116,\ 0.229,\ 0.341$. Such choice of the parameters allowed us to test the validity of our data against (\onlinecite{dickman1997entropic}). In addition we introduced a well (shoulder) big-small interaction of the kind of  \ref{eq:SquareWellPotential}, using $\lambda \sigma_{12} = 3.5,\ 5.5$ (respectively in the case $\xi = 5, 10$) and whose depth (height) $\abs{\epsilon} =  1/\beta = 1$.\\
As for the MC simulations, the size chosen for the box ($H = 22$, $L = 16$ for $\xi = 5$,  $H = 30$, $L = 24$ for $\xi = 10$), was big enough to keep bulk densities within the target values above with a precision of about 1\% in all cases. Runs with different sizes and equilibration lenghts were performed in order to keep size and transient effects under 1\%.\\
Values of the \textit{reduced} force $f_{\rm MC}^*(R) = \beta f_{\rm MC}(R)/(\pi \rho \sigma_{11})$ for values of $R$ ranging from $\sigma_{11}$ to $\sigma_{11} + 3.0\, \sigma_{22}$ at regular intervals of $0.2\, \sigma_{22}$ were obtained. Each data point took about 10-20 hours of CPU time on an Intel Xeon 3.2 GHz processor to be determined. They are shown in \ref{fig:csi5} and \ref{fig:csi10}.
Data taken from (\onlinecite{dickman1997entropic}) (relative to the case $\epsilon = 0$) are also plotted for comparison.

\begin{table}[!ht]
\begin{tabular}{|c|ccc|}
\hline
							& 	$N$ 					&	Sampling frequency (MC steps)	&	Production (MC steps)\\
\hline
$\xi = 5,\ \eta = 0.116$	& 	$\approx 1200$	 		& 	$10^4$	 		& 	$10^{10}$ \\					
$\xi = 5,\ \eta = 0.229$	& 	$\approx 2360$	 		& 	$4 \cdot 10^4$ 	& 	$2 \cdot 10^{10}$ \\	
$\xi = 5,\ \eta = 0.341$	& 	$\approx 3520$	 		& 	$5 \cdot 10^5$	& 	$4 \cdot 10^{10}$ \\
$\xi = 10,\ \eta = 0.116$	& 	$\approx 2700$	 		& 	$10^4$	 		& 	$10^{10}$ \\					
$\xi = 10,\ \eta = 0.229$	& 	$\approx 5340$	 		& 	$4 \cdot 10^4$ 	& 	$2 \cdot 10^{10}$ \\	
$\xi = 10,\ \eta = 0.341$	& 	$\approx 7950$	 		& 	$5 \cdot 10^5$ 	& 	$4 \cdot 10^{10}$ \\
\hline
\end{tabular}
\caption{Values of the parameters used to obtain the $f_{\rm MC}^*(R)$ profiles. By a MC step here we mean a single particle displacement attempt. The averages have been evaluated by using the reported sampling frequency over the total number of MC steps shown in the last column. In all cases at least $10^8$ equilibration MC steps were performed.}
\end{table}

In the PY+S method values of $g_{12}(r)$ were sampled on an equispaced ($dr = 0.01$) mesh of 4096 or 8192 points in $r$-space respectively for $\xi = 5, 10$. Values for any $r$ were obtained from the discrete sample through linear interpolation (linear extrapolation was used to obtain the values near discontinuous points). These in turn allowed us to obtain $\rho_{\rm PY+S}(\vs{r}; R)$ for any $\vs{r}$ through  \ref{eq:SuperpositionApproximation}. The $f_{\rm PY+S}^*(R)$ could finally be obtained performing a numerical integration of the RHS side of  \ref{eq:FinalForce} for various values of $R$. The force profiles are shown in \ref{fig:csi5}, \ref{fig:csi10}. Each required a few minutes of CPU time on a desktop computer. The same procedure was carried out using the HNC+S method, and results are shown in \ref{fig:csi5} for the case $\xi = 5,\ \eta = 0.116$.\\

Forces profiles in the DL approximation are the dotted lines in the \ref{fig:csi5} and \ref{fig:csi10} and were obtained by straightforward substitution of the parameters above in  \ref{eq:DLFinalForce}. \\

The phase diagram on the plane ($\eta_1$, $\eta$) was obtained in the particular case of $\xi = 5$ for different values of $\epsilon_{12}$ and is shown in  \ref{fig:DLPhaseDiagram}.
The associated effective potentials (shown in  \ref{fig:DLPhasePotentials}) were obtained by numerical integration of samples of the force in  \ref{eq:DLFinalForce}. The first order perturbation in in  \ref{eq:FirstOrderFreeEnergy} was also found by numerical integration, while the derivative required by in  \ref{eq:ChemicalPotential} was obtained via numerical differentiation. The discretization of real space needed to carry out such operations was performed on a grid sufficiently fine so that further refinements had no appreciable effect on the scale of the plots.\\

\begin{figure}[!ht]
	\includegraphics[width=.9\textwidth]{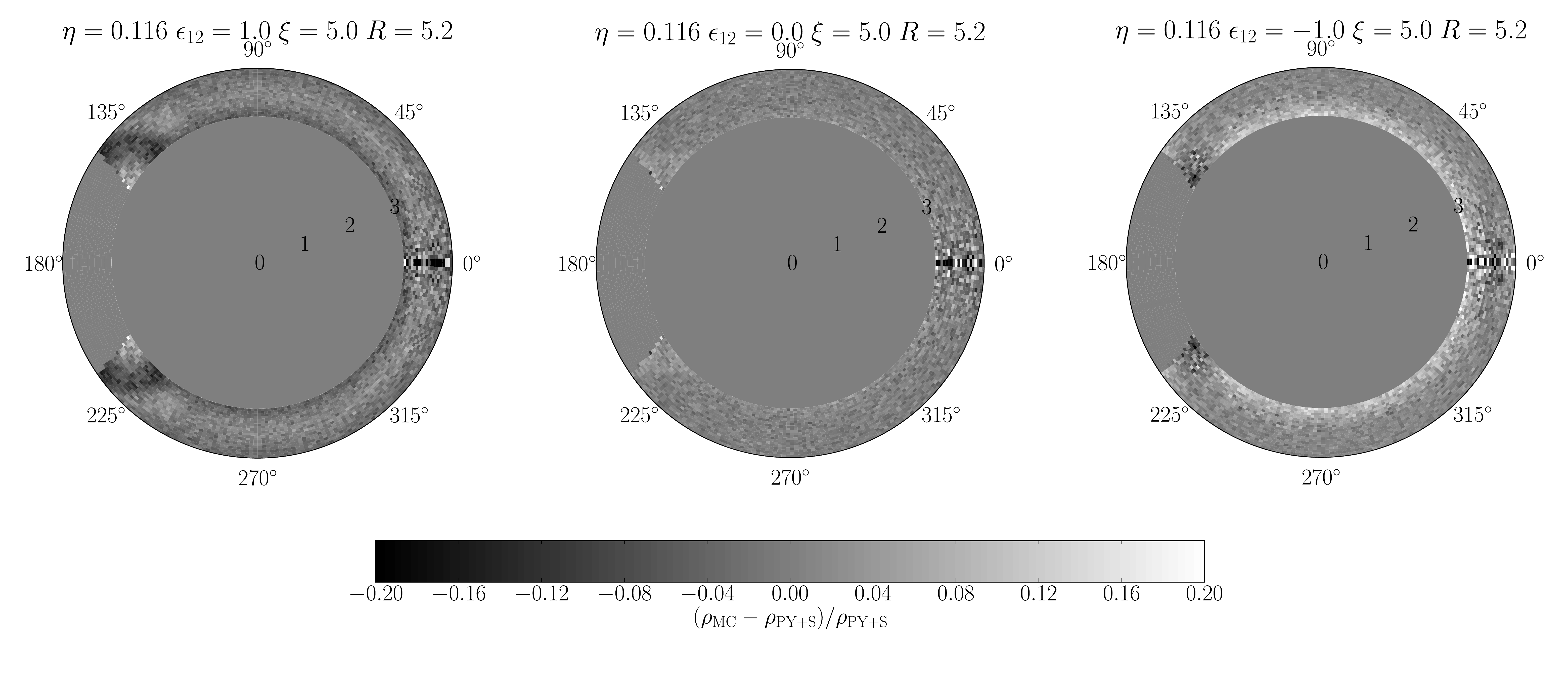}
	\caption{Plot of the difference of the densities of smaller particles around a pair of large particles obtained via MC and PY+S for different values of $\epsilon_{12}$. \label{fig:densityHSSWSS5}}
\end{figure}

In addition we obtained further information about the density of microspheres for the case $\xi = 5$ and $\eta = 0.116$. 2D plots were obtained projecting the MC data in two dimensions exploiting the azimuthal symmetry of the problem. Equivalent diagrams were obtained via the PY+S method. In \ref{fig:densityHSSWSS5}, we show plots of $[\rho_{\rm MC}(r, \theta; R) - \rho_{\rm PY+S}(r, \theta; R)]/ \rho_{\rm PY+S}(r, \theta; R)$ at $R = 5.2$. Such diagrams allow to examine the differences between exact (though noisy) MC data and the approximate PY+S predictions. Such plots are obtained setting $\rho_{\rm MC}(r, \theta; R) \equiv \rho_{\rm MC}(r, - \theta; R)$ when $\theta < 0$, and noise at $\theta \approx 0$ is due to the bad statistics of the particles counts in this region (which is related to the small size of the bins $= 2 \pi r^2 \sin \theta\: dr d\theta$).

\section{\label{sec:Results}Results}

The MC method allows to measure the \textit{exact} effective force in the various cases. Our data are not always in good agreement with those found in (\onlinecite{dickman1997entropic}) for the case $\epsilon_{12} = 0$ and differences up to 20\% are observed. Several simulations with different box sizes and production durations were performed and all our values obtained were consistent between each other within statistical error (which is way lower than 20\%), confirming that our data are reliable.\\
In all cases examined here the DL approximation fails to describe MC data quantitatively. Still its results are in qualitative agreement with the exact profiles which show the features described in \ref{sec:DLapprox}. Better results can be obtained via the PY+S approximation, that describes well the force profiles for $\epsilon_{12} = 0$ and yields a reasonably good agreement also for $\epsilon_{12} = \pm 1$ at low densities and largest distances.\\
Examination of the plots of the differences in densities with the MC and PY+S method elucidates the origin of the differences obtained in the force profiles. The satisfactory result obtained when $\epsilon_{12} = 0$ is mirrored by a quite accurate match between $\rho_{\rm MC}$ and $\rho_{\rm PY+S}$. In this case PY+S only slightly underestimates the particle density in the zone close to both spheres (see \ref{fig:densityHSSWSS5}). The fact that the density match is very good everywhere but in this region suggests that this discrepancy is due to the superposition approximation. 
In the $\epsilon_{12} = \pm 1$ case the density differences are much more pronounced, again in the region close to both the macrospheres (indicating a breakdown of the superposition approximation) but also everywhere else in the vicinity of a single macrosphere (due to loss of accuracy of the PY closure).
Similar results (shown here only in the case $\xi = 5,\ \eta = 0.116$) were obtained via the HNC+S method, confirming that the closure plays a lesser important role than the superposition approximation in the disagreement with the MC force profiles. This last point is evident in the case $\eta = 0.116$ of \ref{fig:csi5}, where the PY+S and HNC+S methods both  fail to describe accurately the force profile at short ranges.\\
The phase diagram (see  \ref{fig:DLPhaseDiagram}) obtained from first order thermodynamic perturbation theory within the DL approximation shows that for small positive values of  $\epsilon_{12}$ one needs to move at higher densities of the smaller species as $\epsilon_{12}$ increases in order to observe phase separation. This is due to the fact that in this regime an increase in $\epsilon_{12}$ corresponds to an additional repulsive term in the effective potential (see  \ref{fig:DLPhasePotentials}).

Such ``stabilizing effect'' due to mutual attraction doesn't hold at higher values of $\epsilon_{12}$, where increments in $\epsilon_{12}$ correspond to progressively stronger effective attraction and lowering of the coexistence line in the $(\eta_1, \eta)$ plane. The critical density in the case under examination is higher in the low $\epsilon_{12}$ regime than in the high $\epsilon_{12}$ one.

\section{\label{sec:Conclusion}Conclusions}

In this paper we have studied the effective forces between two big hard spheres dispersed into a fluid of smaller particles. In particular we studied the effect of interspecies interactions. To this aim we have extended the force derivation done by Attard to the case in which these take the shape of a square well or shoulder, that is in the case where both entropic and energetic effects play a role. As in the case of the original formula, it is sufficient to know the density profile of the small particles around the big ones to obtain the exact value of the effective force. The determination of the density however is a difficult task and one has to rely on some approximations. Here we have proposed an equivalent of the Asakura-Oosawa approximation, i.e. the assumption that the small particles behave like an ideal gas. This approximation leads to an analytical expression that captures, at least qualitatively, the correct behavior as predicted by our MC simulations.  An approach based on integral equations and a superposition approximation improves the results but stills fails in estimating the density in the neighborhood of the two large spheres regardless the choice of the closure used (at least in the cases examined here).
We have shown how the addition of a shoulder or of a well in the interspecies interaction can change qualitatively the well known depletion force profile observed in hard sphere binary mixtures. The presence of a well, for example, causes the onset of a repulsion at short ranges and, in the case of deep wells, of an attraction at intermediate ranges. This happens because when a well is present the larger particles are surrounded by a layer of smaller ones. This layer makes a close contact between the large particles unfavorable, but at the same time if two bigger spheres share part of their surrounding layers the small particles sitting between them act as ``glue'', stabilizing a configuration of intermediate distance between the pair. Such description qualitatively agrees with what has been observed in molecular simulations of binary mixtures of eye-lens proteins \cite{stradner2007new, dorsaz2008colloidal} of hard-core potentials with a Yukawa tail \cite{louis2002effective} and could be confirmed by experimental determinations of the force on colloids interacting with a square shoulder/potential (polymer grafted colloids might be a valid candidate member to this class) using optical tweezers \cite{crocker1999entropic}.\\
The phase diagram as studied with simple thermodynamic perturbation theory within the DL approximation shows that the introduction of shallow well (low $\epsilon_{12}$) has the effect of pushing to higher densities of the smaller component the phase separation. Increasing further $\epsilon_{12}$, however, one reaches the point where deepening the well has the opposite effect.\\
The effect of steric hindrance of the small particles in the purely hard sphere case has already been claimed to be used to stabilize solutions \cite{wasan2003new} and foods \cite{xu1998fat}. It's clear that taking into account the possibility of tuning the interspecies interactions could broaden even more the possible routes for stabilization. Summarizing, we confirm, in agreement with previous work \cite{stradner2007new, dorsaz2008colloidal}, that the mutual attraction could be an extra parameter to play with when tuning the stability of a binary mixture. The present work provides a qualitative and quantitative analysis of the resulting changes.

\acknowledgement
We would like to thank Phil Attard for useful discussions and Simone Belli who wrote with one of us (D.F.) the Fortran implementation of the Gillan algorithm of solution of the IE. We thank also Francesco Varrato and Nicolas Dorsaz for comments on the manuscript. D.F. and G.F. acknowledge support by the Swiss National Science Foundation (grant no. PP$0022\_119006$).

\appendix
\section{\label{sec:ForceDerivationDetails}Derivation of the effective force}

It is shown in (\onlinecite{attard1989spherically}) that the force can be expressed as

\begin{equation}
	f(R) = - \frac{\partial \phi_{11}}{\partial R} - \frac{\partial F_{2}}{\partial R} = - \frac{\partial \phi_{11}}{\partial R}  + \frac{1}{\beta Z_2} \frac{\partial Z_2}{\partial R},
	\label{eq:EffectiveForceScalar}
\end{equation}

where $\phi_{11}$ is the interaction potential between particles of type 1, $F_{2}$ is the free energy of the particles of type 2 that move in the potential generated by the fixed pair of type 1, $Z_{2}$ the partition function associated to it and $\beta = 1/k_B T$.

Following (\onlinecite{attard1989spherically}) we also have that
\begin{equation}
	\frac{\partial Z_2}{\partial R} =	- \int \frac{\partial}{\partial R}\left(1 - e^{-\beta V(\vs{r})} \right)\ e^{\beta V(\vs{r})} \rho (\vs{r}) Z_2 \ \vs{dr},
	\label{eq:DeadlyIntegral}
\end{equation}
where $V(\vs{r})$ is the potential energy of a single particle of type 2 due to the presence of the fixed couple of type 1.  $V(\vs{r})$ can be written as
\begin{equation}
	V(\vs{r}) = \left\{
		\begin{array}{rl}
				\infty				& 	\text{if }	r < \sigma_{12} \text{ or } \abs{\vs{r} - \vs{R}} < \sigma_{12}\\
				- \epsilon_{12}		& 	\text{if }	\sigma_{12} \leq r < \lambda \sigma_{12}  \text{ xor } \sigma_{12} \leq \abs{\vs{r} - \vs{R}} < \lambda \sigma_{12}\\
				-2 \epsilon_{12}	& 	\text{if }	\sigma_{12} \leq r < \lambda \sigma_{12}  \text{ and } \sigma_{12} \leq \abs{\vs{r} - \vs{R}} < \lambda \sigma_{12}\\
				0					& 	\text{otherwise}
		\end{array}
	\right. .
	\label{eq:ExternalPotentialDetail}
\end{equation}

We can write the resulting Mayer function as:
\begin{align}
		1 - e^{-\beta V(\vs{r})}	&= (1 - e^{\beta \epsilon_{12}}) \left[\mathscr{H}_{\lambda \sigma_{12}}(\vs{r}) + \mathscr{H}_{\lambda \sigma_{12}}(\vs{r - R})\right] + \nonumber \\
									&\quad + (2 e^{\beta \epsilon_{12}} - e^{2 \beta \epsilon_{12}} - 1) \mathscr{H}_{\lambda \sigma_{12}}(\vs{r})  \mathscr{H}_{\lambda \sigma_{12}}(\vs{r - R}) + \nonumber \\
									&\quad + (e^{2 \beta \epsilon_{12}} - e^{\beta \epsilon_{12}}) \left[ \mathscr{H}_{\sigma_{12}}(\vs{r})  \mathscr{H}_{\lambda \sigma_{12}}(\vs{r - R}) + \mathscr{H}_{\lambda \sigma_{12}}(\vs{r})  \mathscr{H}_{\sigma_{12}}(\vs{r - R}) \right] + \nonumber \\
									&\quad - e^{2 \beta \epsilon_{12}} \mathscr{H}_{\sigma_{12}}(\vs{r})  \mathscr{H}_{\sigma_{12}}(\vs{r - R}) + \nonumber \\
									&\quad + e^{\beta \epsilon_{12}} \left[ \mathscr{H}_{\sigma_{12}}(\vs{r}) + \mathscr{H}_{\sigma_{12}}(\vs{r - R})\right],
	\label{eq:MyMayerCharacteristicDecomposition}
\end{align}

where the geometric definition of the support of $V(\vs{r})$ has been encoded using the characteristic functions $\mathscr{H}_D(\vs{r})$ defined as:

\begin{equation}
	\mathscr{H}_D(\vs{r}) = \left\{
		\begin{array}{rl}
			1		& \text{if }	r < D\\
			0		& \text{if }	r \geq D\\
		\end{array}
	\right..
\end{equation}

Plugging  \ref{eq:MyMayerCharacteristicDecomposition} inside  \ref{eq:DeadlyIntegral} and rearranging yields:

\begin{align}
	\frac{\partial Z_2}{\partial R} 	&=	- (1 - e^{\beta \epsilon_{12}}) \int \frac{\vs{R}}{R} \cdot \frac{\vs{r - R}}{|\vs{r - R}|}\ \delta (|\vs{r - R}| - \lambda \sigma_{12}) \ e^{\beta V(\vs{r})} \rho (\vs{r}) Z_2 \ \vs{dr} + \nonumber \\
										&\quad	- (2 e^{\beta \epsilon_{12}} - e^{2 \beta \epsilon_{12}} - 1) \int \mathscr{H}_{\lambda \sigma_{12}}(\vs{r})  \frac{\vs{R}}{R} \cdot \frac{\vs{r - R}}{|\vs{r - R}|}\ \delta (|\vs{r - R}| - \lambda \sigma_{12}) \ e^{\beta V(\vs{r})} \rho (\vs{r}) Z_2 \ \vs{dr} + \nonumber \\
										&\quad	- (e^{2 \beta \epsilon_{12}} - e^{\beta \epsilon_{12}}) \int \mathscr{H}_{\sigma_{12}}(\vs{r}) \frac{\vs{R}}{R} \cdot \frac{\vs{r - R}}{|\vs{r - R}|}\ \delta (|\vs{r - R}| - \lambda \sigma_{12}) e^{\beta V(\vs{r})} \rho (\vs{r}) Z_2 \ \vs{dr} + \nonumber \\  
										&\quad  - (e^{2 \beta \epsilon_{12}} - e^{\beta \epsilon_{12}}) \int \mathscr{H}_{\lambda \sigma_{12}}(\vs{r}) \frac{\vs{R}}{R} \cdot \frac{\vs{r - R}}{|\vs{r - R}|}\ \delta (|\vs{r - R}| - \sigma_{12}) e^{\beta V(\vs{r})} \rho (\vs{r}) Z_2 \ \vs{dr} + \nonumber \\
										&\quad	+ e^{2 \beta \epsilon_{12}} \int \mathscr{H}_{\sigma_{12}}(\vs{r}) \frac{\vs{R}}{R} \cdot \frac{\vs{r - R}}{|\vs{r - R}|}\ \delta (|\vs{r - R}| - \sigma_{12}) \ e^{\beta V(\vs{r})} \rho (\vs{r}) Z_2 \ \vs{dr} + \nonumber \\
										&\quad	- e^{\beta \epsilon_{12}} \int \frac{\vs{R}}{R} \cdot \frac{\vs{r - R}}{|\vs{r - R}|}\ \delta (|\vs{r - R}| - \sigma_{12}) \ e^{\beta V(\vs{r})} \rho (\vs{r}) Z_2 \ \vs{dr},
	\label{eq:IntegralsList}
\end{align}


where we have used the shorthand notation $\lambda \sigma_{12}$ to refer to $\lambda_{12} \sigma_{12}$.

The integrals in  \ref{eq:IntegralsList} can be evaluated in the three dimensional space performing the change of variables $\vs{s} \equiv \vs{r - R}$, using spherical coordinates centered in $\vs{R}$ with $\theta$ defined by $\vs{s} \cdot \vs{R} = s R \cos \theta$ and exploiting the azimuthal symmetry of the problem. 

The first one thus reads
\begin{align}
	&\quad \int \frac{\vs{R}}{R} \cdot \frac{\vs{s}}{s}\, \delta (s - \lambda \sigma_{12}) \ e^{\beta V(\vs{s + R})} \rho (\vs{s + R}) Z_2 \, \vs{ds} = \nonumber \\
	&= \int_0^{\infty}\! \int_0^{2 \pi}\! \int_0^{\pi}\! \frac{\vs{R}}{R} \cdot \frac{\vs{s}}{s}\, \delta (s - \lambda \sigma_{12})\, e^{\beta V(\vs{s + R})} \rho (\vs{s + R}) Z_2 \, s^2 \sin{\theta}\ ds\, d\theta\, d\phi = \nonumber \\
	&= 2 \pi \lambda^2 \sigma_{12}^2 \int_0^{\pi} \cos{\theta} e^{\beta V(\vs{S}_{\lambda \sigma_{12}})} \rho (\vs{S}_{\lambda \sigma_{12}}) Z_2 \sin{\theta} d\theta,
\label{eq:FirstIntegral}
\end{align}
where in the last line $\vs{S}_{\lambda \sigma_{12}}$ is defined with the notation 
\begin{equation}
	\vs{S}_{v} \equiv \vs{v + R} \qquad \textrm{with } \abs{\vs v} = v.
	\label{eq:SDefinition}
\end{equation}

The sixth integral in  \ref{eq:IntegralsList} has the same form and its value is
\begin{equation}
	2 \pi \sigma_{12}^2 \int_0^{\pi} \cos{\theta} e^{\beta V(\vs{S}_{\sigma_{12}})} \rho (\vs{S}_{\sigma_{12}}) Z_2 \sin{\theta} d\theta\,
	\label{eq:SixthIntegral}.
\end{equation}

As for the second
\begin{align}
	&\quad \int \mathscr{H}_{\lambda \sigma_{12}}(\vs{r})  \frac{\vs{R}}{R} \cdot \frac{\vs{r - R}}{|\vs{r - R}|}\ \delta (|\vs{r - R}| - \lambda \sigma_{12}) \ e^{\beta V(\vs{r})} \rho (\vs{r}) Z_2 \ d\vs{r} = \nonumber \\
	&= 2 \pi \lambda^2 \sigma_{12}^2 \int_0^{\pi} \mathscr{H}_{\lambda \sigma_{12}}(\vs{S}_{\lambda \sigma_{12}}) \cos{\theta} e^{\beta V(\vs{S}_{\lambda \sigma_{12}})} \rho (\vs{S}_{\lambda \sigma_{12}}) Z_2 \sin{\theta} d\theta\, .	\label{eq:SecondIntegral}
\end{align}
whereas the fourth
\begin{align}
	&\quad \int \mathscr{H}_{\lambda \sigma_{12}}(\vs{r}) \frac{\vs{R}}{R} \cdot \frac{\vs{r - R}}{|\vs{r - R}|}\ \delta (|\vs{r - R}| - \sigma_{12})  \ e^{\beta V(\vs{r})} \rho (\vs{r}) Z_2 \ d\vs{r} = \nonumber \\
	&= 2 \pi \sigma_{12}^2 \int_0^{\pi} \mathscr{H}_{\lambda \sigma_{12}}(\vs{S}_{\sigma_{12}}) \cos{\theta} e^{\beta V(\vs{S}_{\sigma_{12}})} \rho (\vs{S}_{\sigma_{12}}) Z_2 \sin{\theta} d\theta\, .
	\label{eq:FourthIntegral}
\end{align}
while the third and the fifth ones vanish because wherever $\rho$ is nonzero $\mathscr{H}_{\sigma_{12}}$ is zero and viceversa.\\
The $\mathscr{H}$'s can be eliminated introducing the auxiliary functions
\begin{align}
	{}^1 \rho_{\lambda \sigma_{12}}(\vs{S}_{\lambda \sigma_{12}}) &= \left\{
		\begin{array}{rl}
			\rho(\vs{S}_{\lambda \sigma_{12}})		& \text{if }	0 \leq \theta < {}^1 \theta_{\lambda \sigma_{12}} \\
			0						& \text{otherwise}\\
		\end{array}		
	\right. , \nonumber
	\\
	{}^2 \rho_{\lambda \sigma_{12}}(\vs{S}_{\lambda \sigma_{12}}) &= \left\{ 
		\begin{array}{rl}
			\rho(\vs{S}_{\lambda \sigma_{12}})		& \text{if }	{}^1 \theta_{\lambda \sigma_{12}} \leq \theta < {}^2 \theta_{\lambda \sigma_{12}} \leq \pi\\
			0						& \text{otherwise}\\
		\end{array}		
	\right. , \nonumber
	\\
	{}^1 \rho_{\sigma_{12}}(\vs{S}_{\sigma_{12}}) &= \left\{
		\begin{array}{rl}
			\rho(\vs{S}_{\sigma_{12}})	& \text{if }	0 \leq \theta < {}^1 \theta_{\sigma_{12}} \\
			0							& \text{otherwise}\\
		\end{array}		
	\right. ,\nonumber
	\\
	{}^2 \rho_{\sigma_{12}}(\vs{S}_{\sigma_{12}}) &= \left\{
		\begin{array}{rl}
			\rho(\vs{S}_{\sigma_{12}})	& \text{if }	{}^1 \theta_{\sigma_{12}} \leq \theta < {}^2 \theta_{\sigma_{12}} \leq \pi\\
			0						& \text{otherwise}\\
		\end{array}		
	\right. ,
\end{align}
where
\begin{align}
	{}^1 \theta_{\lambda \sigma_{12}} &= \arccos\left(\frac{R^2}{2\, \lambda \sigma_{12}\, R}\right) \nonumber \\
	{}^2 \theta_{\lambda \sigma_{12}} &= \arccos\left(\frac{\lambda^2 \sigma_{12}^2 - R^2 - \sigma_{12}^2}{-2\, \sigma_{12}\, R}\right) \nonumber \\
	{}^1 \theta_{\sigma_{12}} &= \arccos\left(\frac{\sigma_{12}^2 - R^2 - \lambda^2 \sigma_{12}^2}{-2\, \lambda \sigma_{12}\, R}\right) \nonumber \\
	{}^2 \theta_{\sigma_{12}} &= \arccos\left(\frac{R^2}{2\, \sigma_{12}\, R}\right)
\label{eq:Angles}
\end{align}
 are the angles at which the spheres of radii $\sigma_{12}$ and $\lambda \sigma_{12}$ centered in $\vs{0}$ and $\vs{R}$ intersect.
Defining
\begin{align}
	\rho(\vs{S}_{\lambda \sigma_{12}})		&=	{}^1 \rho_{\lambda \sigma_{12}}(\vs{S}_{\lambda \sigma_{12}}) + {}^2 \rho_{\lambda \sigma_{12}}(\vs{S}_{\lambda \sigma_{12}}) \nonumber \\
	\rho(\vs{S}_{\sigma_{12}})	&=	{}^1 \rho_{\sigma_{12}}(\vs{S}_{\sigma_{12}}) + {}^2 \rho_{\sigma_{12}}(\vs{S}_{\sigma_{12}})
	\label{eq:OtherRho}
\end{align}
and substituting these inside the integrals and some algebra eventually yields  \ref{eq:FinalForce}.

\section{\label{sec:nastyDLforce}Expression of the force in the dilute limit}

Use of  \ref{eq:DLapprox} together with  \ref{eq:ExternalPotentialDetail} allows to rewrite the densities that appear in  \ref{eq:FinalForce} as

\begin{align}
	\rho(\vs{S}_{\lambda \sigma_{12}}) &= \left\{
		\begin{array}{rl}
			\rho					& \text{if }	0 \leq \theta < {}^1 \theta_{\lambda \sigma_{12}} \\
			\rho e^{\beta \epsilon_{12}}		& \text{if }	{}^1 \theta_{\lambda \sigma_{12}} \leq \theta < {}^2 \theta_{\lambda \sigma_{12}} \\
			0						& \text{otherwise}\\
		\end{array}		
	\right. , \nonumber
	\\
	\rho(\vs{S}_{\sigma_{12}}) &= \left\{
		\begin{array}{rl}
			\rho e^{\beta \epsilon_{12}}		& \text{if }	0 \leq \theta < {}^1 \theta_{\sigma_{12}} \\
			\rho e^{2 \beta \epsilon_{12}}		& \text{if }	{}^1 \theta_{\sigma_{12}} \leq \theta < {}^2 \theta_{\sigma_{12}} \\			
			0			& \text{otherwise}\\
		\end{array}		
	\right. . \label{eq:RhoInTheDLapprox}
\end{align}
where the angles are the same defined in  \ref{eq:Angles}. Substitution of such density inside \ref{eq:FinalForce} yields a piecewise expression of the force valid for $R \in [\sigma_{11}, + \infty]$:

\fbox{if $\sigma_{11} < R \leq 2\sigma_{12}$}:
\begin{align*}
	f_{\rm DL}(R)	
				&	= -  \frac{\pi \rho}{\beta}			\left\{ 
											e^{\beta \epsilon_{12}} \left(
												- \left(\frac{R^2 + \sigma_{12}^2 -\lambda^2 \sigma_{12}^2}{2 R}\right)^2 + \sigma_{12}^2 
											\right) +
										\right. \nonumber \\   
				&	\qquad			
										\left.
											+ e^{2 \beta \epsilon_{12}} \left(
												- \left(\frac{R}{2}\right)^2 + \left(\frac{R^2 + \sigma_{12}^2 -\lambda^2 \sigma_{12}^2}{2 R}\right)^2
											\right) +
										\right. \nonumber \\
				&	\qquad			
										\left.
											+ (1 - e^{\beta \epsilon_{12}}) \left[ 
												\left(
												- \left(\frac{R}{2}	\right)^2 + \lambda^2 \sigma_{12}^2 
												\right)
											\right. +
										\right. \nonumber \\
				&	\qquad				
										\left.
											\left.
												\qquad \qquad + e^{\beta \epsilon_{12}} \left(
													- \left(\frac{R^2 + \lambda^2 \sigma_{12}^2 - \sigma_{12}^2}{2 R}\right)^2 + \left(\frac{R}{2}	\right)^2
												\right)
											\right]					
										\right\};
\end{align*}

\fbox{if $2 \sigma_{12} < R \leq \sigma_{12} + \lambda \sigma_{12}$}:
\begin{align*}
	f_{\rm DL}(R) 	& = - \frac{\pi \rho}{\beta} 		\left\{ 
											e^{\beta \epsilon_{12}} \left(
												- \left(\frac{R^2 + \sigma_{12}^2 -\lambda^2 \sigma_{12}^2}{2 R}\right)^2 + \sigma_{12}^2 
											\right) +
										\right. \nonumber \\   
				&	\qquad			
										\left.
											+ e^{2 \beta \epsilon_{12}} \left(
												- \sigma_{12}^2 + \left(\frac{R^2 + \sigma_{12}^2 -\lambda^2 \sigma_{12}^2}{2 R}\right)^2
											\right) +
										\right. \nonumber \\
				&	\qquad			
										\left.
											+ (1 - e^{\beta \epsilon_{12}}) \left[
												\left(
												- \left(\frac{R}{2}	\right)^2 + \lambda^2 \sigma_{12}^2 
												\right)
											\right. +
										\right. \nonumber \\
				&	\qquad				
										\left.
											\left.
												\qquad \qquad + e^{\beta \epsilon_{12}} \left(
													- \left(\frac{R^2 + \lambda^2 \sigma_{12}^2 - \sigma_{12}^2}{2 R}\right)^2 + \left(\frac{R}{2}	\right)^2
												\right)
											\right]					
										\right\};
\end{align*}

\fbox{if $\sigma_{12} + \lambda \sigma_{12} < R \leq 2 \lambda \sigma_{12}$}:
\begin{align*}
	f_{\rm DL}(R) 	&= - \frac{\pi \rho}{\beta}	\left\{
											(1 - e^{\beta \epsilon_{12}}) \left[ 
												\left(
												- \left(\frac{R}{2}	\right)^2 + \lambda^2 \sigma_{12}^2 
												\right)
											\right. +
										\right. \nonumber \\
				&	\qquad				
										\qquad \left.
											\left.
												\qquad + e^{\beta \epsilon_{12}} \left(
													- \lambda^2 \sigma_{12}^2 + \left(\frac{R}{2}	\right)^2
												\right)
											\right]					
										\right\};
\end{align*}

\fbox{if $2 \lambda \sigma_{12} < R$}:
\begin{align}
	f_{\rm DL}(R) 	& = 0 . \label{eq:DLFinalForce} 
\end{align}

\section{Linear dependence of volume terms with $\rho_1$ \label{sec:VolumeTerms}}
\begin{figure}[!ht]
	\subfigure[\label{fig:DLPhasePotentials}]{\includegraphics[width=.48\textwidth]{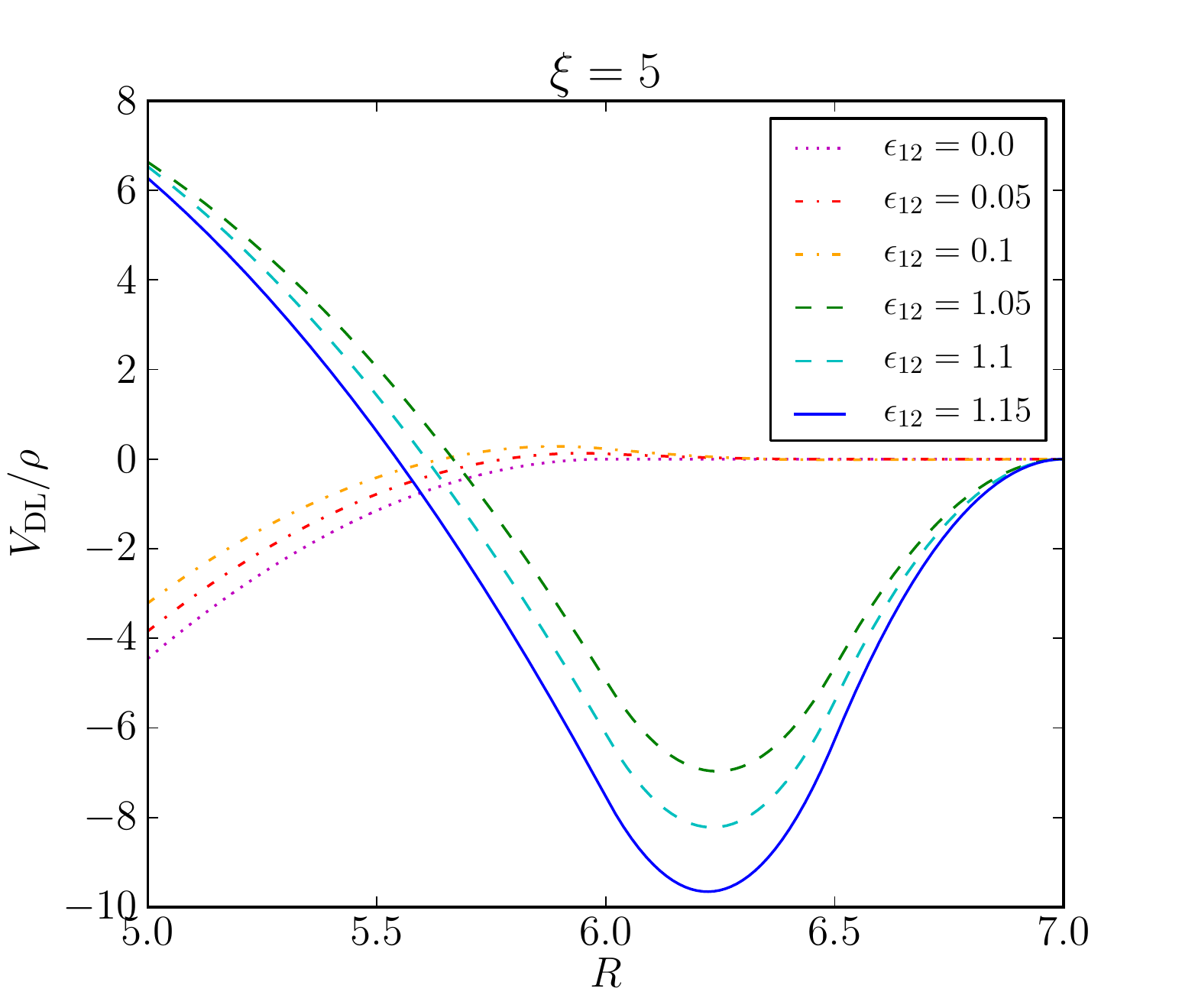}}\\
	\subfigure[\label{fig:DLPhaseDiagram}]{\includegraphics[width=.48\textwidth]{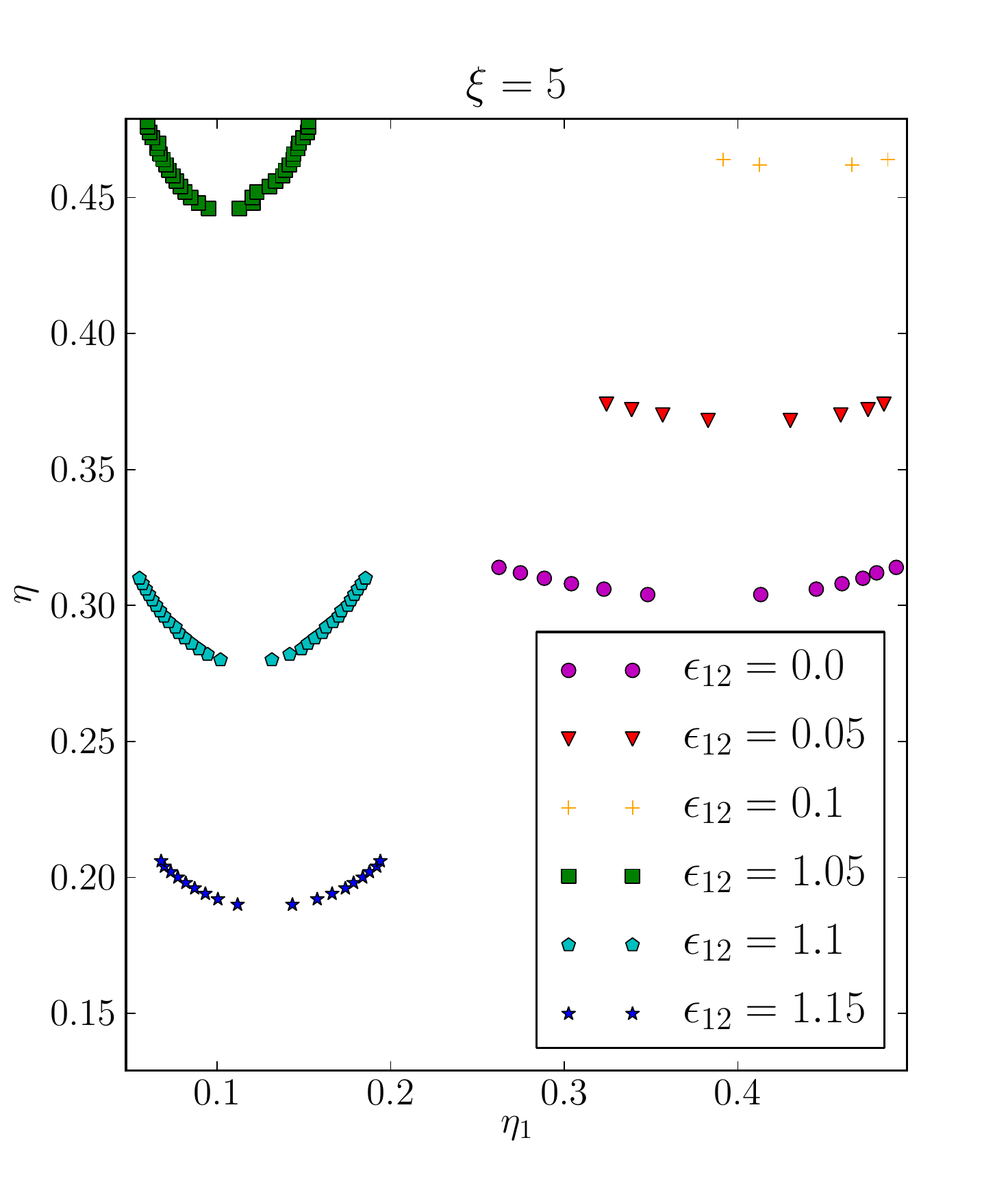}}
	\caption{
	\ref{fig:DLPhasePotentials}: Effective potentials obtained by numerical integration of the force in \ref{eq:DLFinalForce} using the parameters reported in the text relative to the case $\xi = 5$ for various values of $\epsilon_{12}$. 
	\ref{fig:DLPhaseDiagram}: Coexistence curves obtained by first order perturbation theory using the potentials in \ref{fig:DLPhasePotentials}.
	\label{fig:DLPhase}}
\end{figure}

Following Dijkstra et al. \cite{dijkstra1999phase} we calculate the volume terms (0- and 1-body) of $\Omega$ in the effective potential of \ref{eq:EffectivePotential}.\\
The first term can be interpreted as the grand potential of a pure system of small particles at fugacity $z_2$ enclosed in a volume $V$
\begin{equation}
	\Omega_0 = \frac{z_2}{\beta} \int_V \vs{dr} = \frac{z_2 V}{\beta}, \label{eq:0BodyTerm}
\end{equation}
while the second is
\begin{align}
	\Omega_1 	&= \sum_N \frac{z_2}{\beta} \int_V f_i\ \vs{dr} \\
				&= \sum_N \frac{z_2}{\beta} \left(\frac{4\pi}{3} \sigma_{12}^3 ((e^{- \beta \epsilon_{12}} - 1)(\lambda^3 - 1) - 1) \right) \\
				&= \rho_1 V \frac{z_2}{\beta} \times {\rm const}, \label{eq:1BodyTerm}
\end{align}
where we have used the definition $f_i = e^{- \beta \phi_{12}(\vs{R_i - r})} - 1$.
From   \ref{eq:0BodyTerm} and \ref{eq:1BodyTerm} we see that $(\Omega_0 + \Omega_1)/V$ is linear with respect to $\rho_1$ and thus does not alter the results of the phase diagram construction that leads to  \ref{fig:DLPhasePotentials} \cite{likos2001effective, dijkstra1999phase}.

\bibliography{manuscript_Fiocco_etal}

\clearpage

\end{document}